%% file: HH_FTapp_main.tex
\documentclass[12pt]{article}
\usepackage{amsmath} 
\usepackage{longtable}
\usepackage{graphicx,epsfig}    
\usepackage{color}
\usepackage{cite}
\usepackage{url}
\usepackage{amssymb}
\usepackage{xspace}

\newcommand\as{\alpha_{\mathrm{S}}} 
\newcommand\f[2]{\frac{#1}{#2}}

\def\beq{\begin{equation}} 
\def\eeq{\end{equation}} 
\def\to{\rightarrow}

\def\b0{\beta_0}

\def\beeq{\begin{eqnarray}}
\def\eeeq{\end{eqnarray}}

\def\mur{\mu_R} 
\def\muf{\mu_F}
\def\mur2{\mu_R^2} 
\def\muf2{\mu_F^2}

\def\to{\rightarrow}

\newcommand{\sqrtS}{\ensuremath{\sqrt{s}}}

\newcommand\OpenLoops{{\sc OpenLoops}\xspace}
\newcommand\Collier{{\sc Collier}\xspace}

\newcommand{\D}{\mathrm{d}}

\newcommand{\ftapprox}{FT$_{\mathrm{approx}}$\xspace}
\newcommand{\nnloBP}{NNLO$_{\mathrm{B-proj}}$\xspace}
\newcommand{\nnloNI}{NNLO$_{\mathrm{NLO-i}}$\xspace}
\newcommand{\nnloFT}{NNLO$_{\mathrm{FTapprox}}$\xspace}

\usepackage[overlay,absolute]{textpos}
\newcommand\PlaceText[3]{%
\begin{textblock*}{10in}(#1,#2)  
#3
\end{textblock*}
}%

\newcommand {\apgt} {\ {\raise-.5ex\hbox{$\buildrel>\over\sim$}}\ }
\newcommand {\aplt} {\ {\raise-.5ex\hbox{$\buildrel<\over\sim$}}\ }

\def\citere#1{\mbox{Ref.~\cite{#1}}}
\def\citeres#1{\mbox{Refs.~\cite{#1}}}
\def\reffi#1{\mbox{Fig.~\ref{#1}}}
\def\reffitwo#1#2{\mbox{Figs.~\ref{#1} and \ref{#2}}}

\def\refta#1{\mbox{Table~\ref{#1}}}
\def\refse#1{\mbox{Section~\ref{#1}}}

\newcommand\Tstrut{\rule{0pt}{3.0ex}}         
\newcommand\Bstrut{\rule[-1.5ex]{0pt}{0pt}}   

\title{Higgs boson pair production at NNLO with \\[1ex]
top quark mass effects}

\author{Massimiliano Grazzini$^1$, Gudrun Heinrich$^2$, Stephen Jones$^2$, Stefan Kallweit$^3$, 
\\ Matthias Kerner$^2$, Jonas M. Lindert$^4$, Javier Mazzitelli$^1$
\\\\
\footnotesize{$^1$Physik-Institut, Universit\"at Z\"urich, Winterthurerstrasse 190, CH-8057 Z\"urich, Switzerland}
\\
\footnotesize{$^2$Max Planck Institute for Physics, F\"ohringer Ring 6, 80805 M\"unchen, Germany}
\\
\footnotesize{$^3$Theoretical Physics Department, CERN, CH-1211 Geneva 23, Switzerland}
\\
\footnotesize{$^4$Institute for Particle Physics Phenomenology, University of Durham, Durham,~DH1~3LE, UK}
}            

\date{}

\begin{document}
\renewcommand{\thefootnote}{\fnsymbol{footnote}}
\pagenumbering{gobble}

\PlaceText{160mm}{10mm}{
\noindent \small
CERN-TH-2018-044\\
IPPP/18/15 \\
MPP-2018-30\\
ZU-TH 10/18
}

\maketitle

\begin{abstract}
\noindent We consider QCD radiative corrections to Higgs boson pair production through gluon fusion in proton collisions.
We combine the exact next-to-leading order (NLO) contribution, which features two-loop virtual
amplitudes with the full dependence on the top quark mass $M_t$, with the next-to-next-to-leading order (NNLO) corrections computed
in the large-$M_t$ approximation. The latter are improved with different reweighting
techniques in order to account for finite-$M_t$ effects beyond NLO. Our reference NNLO result is obtained by
combining one-loop double-real corrections with full $M_t$ dependence with suitably reweighted real--virtual and double-virtual contributions evaluated in the large-$M_t$ approximation.
We present predictions for inclusive cross sections in $pp$ collisions
at $\sqrt{s}=13$, $14$, $27$ and $100\,$TeV and we discuss their uncertainties
due to missing $M_t$ effects.
Our approximated NNLO corrections increase the NLO result by an amount ranging from +12\% at $\sqrt{s}=13$\,TeV to +7\% at $\sqrt{s}=100$\,TeV, and the residual uncertainty
from missing $M_t$ effects is estimated to be at the few percent level.
Our calculation is fully differential in the Higgs boson pair and the associated jet activity: we also present predictions for various differential distributions at $\sqrt{s}=14$ and $100\,$TeV. 
Our results represent the most advanced perturbative prediction available to date for this process.

\end{abstract}

\newpage

\pagenumbering{arabic}
\section{Introduction}

\input{intro}

\section{Details on the method and approximations}
\label{sec:method}

\input{method}

\section{Results}
\label{sec:results}

\input{results}

\section{Summary}
\label{sec:conclusions}

\input{conclusions}

\subsection*{Acknowledgements}
We thank Stefano Catani, Daniel de Florian, Ramona Gr{\"o}ber, Andreas Maier and Stefano Pozzorini for valuable discussions and comments on the manuscript.
This research was supported in part by the Swiss National Science
Foundation (SNF) under contracts CRSII2-141847, 200021-156585,
and by the Research Executive Agency (REA) of the European
Union under the Grant Agreement number PITN--GA--2012--316704 ({\it HiggsTools}).


\bibliography{biblio}

\end{document}

%% file: intro.tex
One of the primary goals of the LHC programme in the next decades is the detailed study of Higgs boson properties. In particular, the high-luminosity upgrade of the LHC is expected to provide direct constraints on the Higgs boson trilinear coupling from Higgs boson pair production~\cite{CMS-PAS-FTR-16-002,ATL-PHYS-PUB-2017-001}, which may reveal whether the Higgs potential is indeed Standard Model-like. A detailed theoretical understanding of Higgs boson pair production processes is thus mandatory.

Considering the magnitude of the total Higgs boson pair production cross sections at $\sqrt{s}=14$\,TeV~\cite{Baglio:2012np,Frederix:2014hta}, 
the most promising process to constrain the Higgs trilinear  coupling is pair production via gluon fusion. 
Due to the smallness of the corresponding production cross sections, it has been  recently suggested to additionally harness complementary information on the trilinear Higgs coupling from higher-order contributions to single Higgs boson production~\cite{McCullough:2013rea,Gorbahn:2016uoy,Degrassi:2016wml,Bizon:2016wgr,DiVita:2017eyz,Maltoni:2017ims} or electroweak precision observables~\cite{Degrassi:2017ucl,Kribs:2017znd}.

For the $gg\to hh$ production channel, 
the leading order~(LO) calculation was performed some time ago in
\citeres{Eboli:1987dy,Glover:1987nx,Plehn:1996wb}. 
The next-to-leading-order~(NLO) corrections with full top quark mass~($M_t$) dependence, involving two-loop diagrams with several mass scales, 
became available only recently~\cite{Borowka:2016ehy,Borowka:2016ypz}, 
and have been supplemented by soft-gluon resummation at small transverse momenta of the Higgs boson pair~\cite{Ferrera:2016prr}
and parton shower effects~\cite{Heinrich:2017kxx,Jones:2017giv}.

In the $M_t\to\infty$ limit, also called Higgs Effective Field Theory~(HEFT) approximation, 
point-like effective couplings of gluons to Higgs bosons arise. 
In this limit, the NLO corrections were first calculated in 
\citere{Dawson:1998py} and rescaled by a factor $B_{\rm FT}/B_{\rm HEFT}$, where $B_{\rm FT}$
denotes the LO one-loop matrix element squared in the full theory.
This procedure is often called  ``Born-improved HEFT'' approximation.

In \citeres{Frederix:2014hta,Maltoni:2014eza} an approximation for Higgs boson pair production at NLO, labelled
``\ftapprox'', was introduced, in which the real radiation matrix elements 
contain the full top quark mass dependence, while the virtual part is
calculated at NLO in the HEFT approximation and rescaled at the event level by
the re-weighting factor $B_{\rm FT}/B_{\rm HEFT}$. At the inlusive cross section level this approximation suggests at the LHC a correction with respect to the ``Born-improved HEFT'' approximation of about $-10\%$, close to the corresponding correction of $-14\%$ later obtained in the full NLO calculation~\cite{Borowka:2016ehy,Borowka:2016ypz}.

The next-to-next-to-leading-order~(NNLO) QCD corrections in the HEFT approximation have been computed in
\citeres{deFlorian:2013uza,deFlorian:2013jea,Grigo:2014jma,deFlorian:2016uhr},
where \citere{deFlorian:2016uhr} provides fully differential results. 
The NNLO HEFT results for the total cross section have been supplemented by an expansion in $1/M_t^2$ in
\citere{Grigo:2015dia}. Approximations for the top-quark mass dependence of the two-loop amplitudes in the NLO calculation have been studied in \citere{Grober:2017uho} via a Pad\'e ansatz.
Soft gluon resummation has been performed at NLO+NNLL in
\citere{Shao:2013bz} and at NNLO+NNLL in \citere{deFlorian:2015moa}. 
The NNLO+NNLL HEFT results lead to $K$-factors of about 1.2 relative to the Born-improved NLO HEFT result.

In \citere{deFlorian:2016spz}, the recommended value for the total $gg\to hh$ cross section was based on the 
NNLO+NNLL HEFT results~\cite{deFlorian:2015moa}, corrected by a factor $\delta_t$ accounting for top quark mass effects, 
extracted from \citere{Borowka:2016ehy}.
However, this procedure is somewhat ad hoc, and not viable to study kinematical distributions.
In order to account for the NNLO $K$-factor in the HEFT calculation as well as for the correct description of the 
$t\bar{t}$ threshold and the high-energy tails of the distributions, where the top quark loops are resolved, 
a first attempt to combine the two calculations has been made in \citere{Borowka:2016ypz}, 
where the full NLO result for a particular distribution was reweighted by the NNLO $K$-factor obtained from 
\citere{deFlorian:2016uhr} on a bin-by-bin basis. However, this procedure, called ``NLO-improved NNLO''  has its drawbacks, 
as it needs to be repeated for each observable (and binning) under consideration.

The aim of this paper is to study alternative methods to combine the two results, i.e. to incorporate top quark mass effects 
in the calculation of the production of Higgs boson pairs at NNLO. One of the studied approximations comprises exact top-quark mass dependence up to NLO and also exact top quark mass dependence in the double-real emission contributions to the NNLO cross section
at differential level.
The results of this approximation can be regarded as the most advanced prediction currently available for Higgs boson pair production in gluon fusion.

This work is organized as follows: In \refse{sec:method} we describe the technical details of our calculation, and present the different approximations we will consider to incorporate mass effects in the NNLO contribution.
In \refse{sec:results} we present our numerical predictions, both for the total cross section and differential distributions.
Finally, in \refse{sec:conclusions} we summarise our results.

%% file: method.tex
We start by presenting the different technical ingredients entering our computation, as well as the definition of the various approximate ways to include mass effects in the NNLO calculation introduced and used in this work. Finally, we also discuss the numerical stability of our predictions.

\subsection{Technical ingredients}

Our calculation is based on the publicly available computational framework {\sc Matrix}~\cite{Grazzini:2017mhc}, which allows the user to perform fully differential NNLO calculations for a wide class of processes at hadron colliders.
For the purpose of the present work, the public version of the code has been extended, based on the calculation of \citere{deFlorian:2016uhr},
to include the production of a pair of Higgs bosons via gluon fusion.
For the calculation of the NNLO corrections the code implements the $q_T$-subtraction formalism~\cite{Catani:2007vq}, in which the genuine NNLO singularities, located where the transverse momentum of the Higgs boson pair, $p_{T,hh} \equiv q_T$, vanishes, are explicitly separated from the NLO-like singularities in the $hh+\text{jet}$ contribution.
The $q_T$ subtraction formula reads
\beq\label{eq:qTsubtraction}
\D{\sigma}^{hh}_{\mathrm{NNLO}}={\cal H}^{hh}_{\mathrm{NNLO}}\otimes \D{\sigma}^{hh}_{\mathrm{LO}}+\left[ \D{\sigma}^{hh+\text{jet}}_{\mathrm{NLO}}-\D{\sigma}^{\mathrm{CT}}_{\mathrm{NNLO}}\right],
\eeq
where in particular the contribution $\D{\sigma}^{hh+\text{jet}}_{\mathrm{NLO}}$ can be evaluated using any available NLO subtraction procedure
to handle and cancel the corresponding infrared (IR) divergencies\footnote{{\sc Matrix} uses the automated implementation of the Catani-Seymour dipole subtraction method~\cite{Catani:1996jh,Catani:1996vz} within
the Monte Carlo program {\sc Munich}~\cite{munich}.}.
The remaining $q_T \to 0$ divergence is canceled by the process-independent counterterm $\D{\sigma}^{\mathrm{CT}}_{\mathrm{NNLO}}$.
The process-dependence of the hard-collinear coefficient ${\cal H}^{hh}_{\mathrm{NNLO}}$ enters only via the NNLO (HEFT) two-loop virtual corrections~\cite{deFlorian:2013uza} through an appropriate subtraction procedure~\cite{Catani:2013tia}.

The difference in the square bracket of Eq.~(\ref{eq:qTsubtraction}) is finite when $q_T \to 0$, but each of the terms exhibits a logarithmic divergence. Therefore, a technical cut, $r_\text{cut}$, needs to be introduced on $q_T/Q$, where the scale $Q$ is chosen to be the invariant mass of the final-state system.
More details about the $r_\text{cut} \to 0$ extrapolation are provided in \refse{sec:stability}.

At variance with the calculation of \citere{deFlorian:2016uhr}, which was strictly done within the HEFT, this time all the routines needed to compute the full NLO cross section as well as the different NNLO reweightings have been implemented. This includes linking the code to the NLO two-loop virtual corrections obtained via a grid interpolation~\cite{Heinrich:2017kxx} and to several loop-induced amplitudes provided by the {\sc OpenLoops} amplitude generator~\cite{OLhepforge}. Within this framework we reproduced the differential NLO results of \citeres{Borowka:2016ehy,Borowka:2016ypz} at the per mille level.

The grid for the NLO virtual two-loop amplitudes is based on the calculation presented in \citeres{Borowka:2016ehy,Borowka:2016ypz}, which in turn for the calculation of the two-loop amplitudes relies on an extension of the program {\sc GoSam}~\cite{Cullen:2011ac,Cullen:2014yla} to two loops~\cite{Jones:2016bci}, using also~{\sc Reduze2} \cite{vonManteuffel:2012np}, {\sc SecDec3}~\cite{Borowka:2015mxa} and the Quasi-Monte Carlo technique as described in \citere{Li:2015foa} for the numerical integration.
These amplitudes (for fixed values of the Higgs boson and top quark masses) are provided in a two-dimensional grid together with an interpolation framework, which allows us to evaluate them at any phase space point without having to perform the computationally costly two-loop integration.
For more details, see \citeres{Heinrich:2017kxx,gridURL}.

All tree and one-loop amplitudes in the HEFT and also all loop-squared amplitudes in the full theory as discussed below are obtained via a process independent interface to \OpenLoops~\cite{Cascioli:2011va,OLhepforge,Buccioni:2017yxi}.
For the latter this comprises loop-squared amplitudes for $pp \to hh+1,2$ jets, that need to be evaluated in IR divergent unresolved limits. In particular the limit $q_T \to 0$  represents a significant challenge for the numerical stability of the
$hh+2$ jets amplitudes in the full theory. Thanks to the employed algorithms the numerical stability is under control, as discussed in detail in \refse{sec:stability}.
A major element of this stability originates from the employed tensor integral reduction 
library {\tt COLLIER}~\cite{Denner:2016kdg}.

\subsection{Approximations for top-mass effects at NNLO}
\label{sec:approximations}

In the following we present three approximations for the NNLO Higgs boson pair production cross section, which take into account  finite top quark mass effects in different ways.
In all cases, we always include the full NLO result when computing the NNLO prediction, and only apply the different approximations to the ${\cal O}(\as^4)$ contribution.

\paragraph{NNLO$_\text{NLO-i}$}
The NLO-improved NNLO approximation (NNLO$_\text{NLO-i}$) has already been presented in \citere{Borowka:2016ypz}. It can be constructed based on an observable-level multiplicative approach.
In this approximation, for each bin of each histogram we multiply the full NLO result by the ratio between the HEFT NNLO and NLO predictions for this bin.

\paragraph{NNLO$_\text{B-proj}$}
A different approximation can be obtained by reweighting each NNLO event by the ratio of the full and HEFT Born squared amplitudes. We denote this procedure as Born-projected approximation (NNLO$_\text{B-proj}$).
Of course, in order to do so and due to the different multiplicities involved, an appropriate projection to Born-like kinematics is needed; for this purpose we make use of the $q_T$-recoil procedure defined in \citere{Catani:2015vma}.
Following this prescription, the momenta of the Higgs bosons remain unchanged, and the new initial-state parton momenta are obtained by absorbing the recoil due to the additional radiation.
Specifically, denoting the momenta of the incoming partons by $p_1$ and $p_2$, and the momentum of the Higgs boson pair system by $q$, the new momentum to be used for the LO projection $k_1$ (then, $k_2 = q - k_1$) is given by
\beq
k_1^\mu = z_1 \f{Q^2}{2\, q \cdot p_1}\, p_1^\mu
 + k_{1T}^\mu 
 + \f{k_{1T}^2}{z_1} \f{q \cdot p_1}{Q^2 \, p_1 \cdot p_2}\, p_2^\mu \,,
 \hspace*{1cm}
 (k_{1T}^\mu k_{1T\mu} = -k_{1T}^2) \,,
\eeq 
where
\beq
z_1 = \f{Q^2 + 2 \,q_T \cdot k_{1T} + \sqrt{(Q^2 + 2 \,q_T\cdot k_{1T})^2 - 4 Q_T^2 k_{1T}^2}}{2\, Q^2}\,,
\hspace*{1cm}
(Q_T^2 \equiv Q^2 + q_T^2) \,,
\eeq
and $k_{1T}^\mu$ is a two-dimensional vector in the $q_T$ plane which needs to fulfill the condition $k_{1T} \to 0$ when $q_T \to 0$, and we set  $k_{1T} = q_T/2$ (and therefore $k_{2T} = q_T/2$).
This condition guarantees that the subsequently applied reweighting does not spoil the NNLO $q_T$-cancellation. 
More details about this procedure can be found in \citere{Catani:2015vma}.

\paragraph{NNLO$_{\text{FT}_\text{approx}}$}
The third approximation we consider is constructed to profit from the fact that the double-real emission contributions to the NNLO cross section
require only one-loop amplitudes in the full theory~(FT)
and can thus be computed by using \OpenLoops.
Of course, the inclusion of these loop-induced amplitudes needs to be done in such a way that the dipole cancellations in the NLO $hh+j$ calculation and the low-$q_T$ cancellation for $hh$ at NNLO are not spoiled.

We will define our approximation by using the following procedure: working in the HEFT, for each $n$-loop squared amplitude that needs to be computed for a given partonic subprocess ${\cal A}^{(n)}_\text{HEFT}(ij \to HH + X)$, we apply the reweighting
\beq\label{eq:reweight}
{\cal R}(ij \to HH + X) =
\f{{\cal A}^\text{Born}_\text{Full}(ij \to HH + X)}{{\cal A}^{(0)}_\text{HEFT}(ij \to HH + X)}\,,
\eeq
where ${\cal A}^\text{Born}_\text{Full}$ stands for the lowest order (loop-induced) squared amplitude for the corresponding partonic subprocess, computed in the full theory.\footnote{Strictly speaking, the reweighting is applied to the finite part of the loop amplitudes. However, at one-loop level this procedure reproduces the loop structure of the full theory.}
We note that, contrary to what happens in the Born-projected approach, here the reweighting is defined using amplitudes that correspond to the same subprocess under consideration.
Therefore, the kinematics is always preserved and there is no need to define a Born projection.
Moreover, for amplitudes that are of tree-level type in the HEFT (as it is the case for the double-real emission contributions), this reweighting simply implies using the exact loop-induced amplitudes with full top mass dependence.
The reweighting procedure defined by Eq.~(\ref{eq:reweight}) agrees at NLO with the so-called FT$_\text{approx}$ introduced in \citere{Maltoni:2014eza}, therefore we will use the same notation.

Given that the performance of the Born-projection and FT approximations was already studied in Ref.~\cite{Borowka:2016ypz} at NLO, we directly present NNLO predictions in \refse{sec:results}.
We point out that, based on the ingredients entering each of the approximations, the NNLO$_{\text{FT}_\text{approx}}$ is expected to be the most advanced prediction for Higgs boson pair production via gluon fusion.
By contrast, the NNLO$_\text{B-proj}$ is expected to be the less accurate, since it is based on a simple Born level reweighting procedure.
Nevertheless, and for comparison purposes, we always present results for the three approximations described above.

\subsection{Numerical stability}\label{sec:stability}

Before presenting our quantitative predictions, we briefly discuss the numerical stability of our results.
From the computational point of view, the most challenging of the three approaches to incorporate mass effects
at NNLO is the NNLO$_\text{FTapprox}$ procedure, as it involves loop-induced double-real contributions in the full theory. In particular the dominant $gg \to hh gg$ amplitude comprises computationally very challenging six-point loop integrals with internal masses.
In fact, these contributions have to be evaluated in the numerically	intricate
NNLO unresolved limits and to the best of our knowledge, the present calculation is the first application of a six-point one-loop amplitude integrated over its IR divergent unresolved limits in an NNLO calculation.

Thanks to the numerical stability of the applied algorithms in \OpenLoops together with \Collier, the bulk of the phase-space points remains stable in double precision
when approaching $q_T \to 0$, even close to the dipole singularity, i.e. in the NNLO double-unresolved limits. On average the runtime per phase space point for the $gg \to hh gg$ amplitude is $\sim 1$\,sec.
In principle \OpenLoops provides a rescue system, such that remaining numerically unstable phase-space points can be reevaluated in higher numerical precision based on reduction with {\sc CutTools}~\cite{Ossola:2007ax}. However, the runtime of the loop-induced $gg\to hh gg$ amplitude in \OpenLoops is significantly increased when {\sc CutTools} is used in quadruple precision (to the level of $\sim10$ minutes per phase-space point), rendering the quadruple precision stability system prohibitive for this amplitude for
practical purposes\footnote{Here we want to note that these stability issues will be strongly mitigated in the future based on the new \OpenLoops on-the-fly reduction method introduced in \citere{Buccioni:2017yxi}.}.
Therefore, we restrict the evaluation to double precision and replace potentially unstable phase-space points close to the dipole singularities,
quantified by $\alpha_\text{L-i} = (p_i \cdot p_j/\hat s)_\text{min}$, where the minimum among all potential emitter parton combinations $i$ and $j$ is taken,
with an approximation: Below a technical cut 
$\alpha_\text{L-i, \rm{cut}}$ we switch from the (loop-induced) double-real amplitude in the FT to the (tree-level) double-real amplitudes in the HEFT, reweighted at LO.
This approach could in principle introduce a bias in the NLO $hh+$jet cross section, thereby hampering the low-$q_T$ cancellation of the NNLO computation. We have checked that this is not the case, as detailed in the following.

For the predictions presented in \refse{sec:results} we use $\alpha_\text{L-i, \rm{cut}} = 10^{-4}$ and we varied this parameter in the range $10^{-3}$ to $10^{-5}$, finding independence of all results. 
%
\begin{figure}
\begin{center}
\includegraphics[width=.49\textwidth]{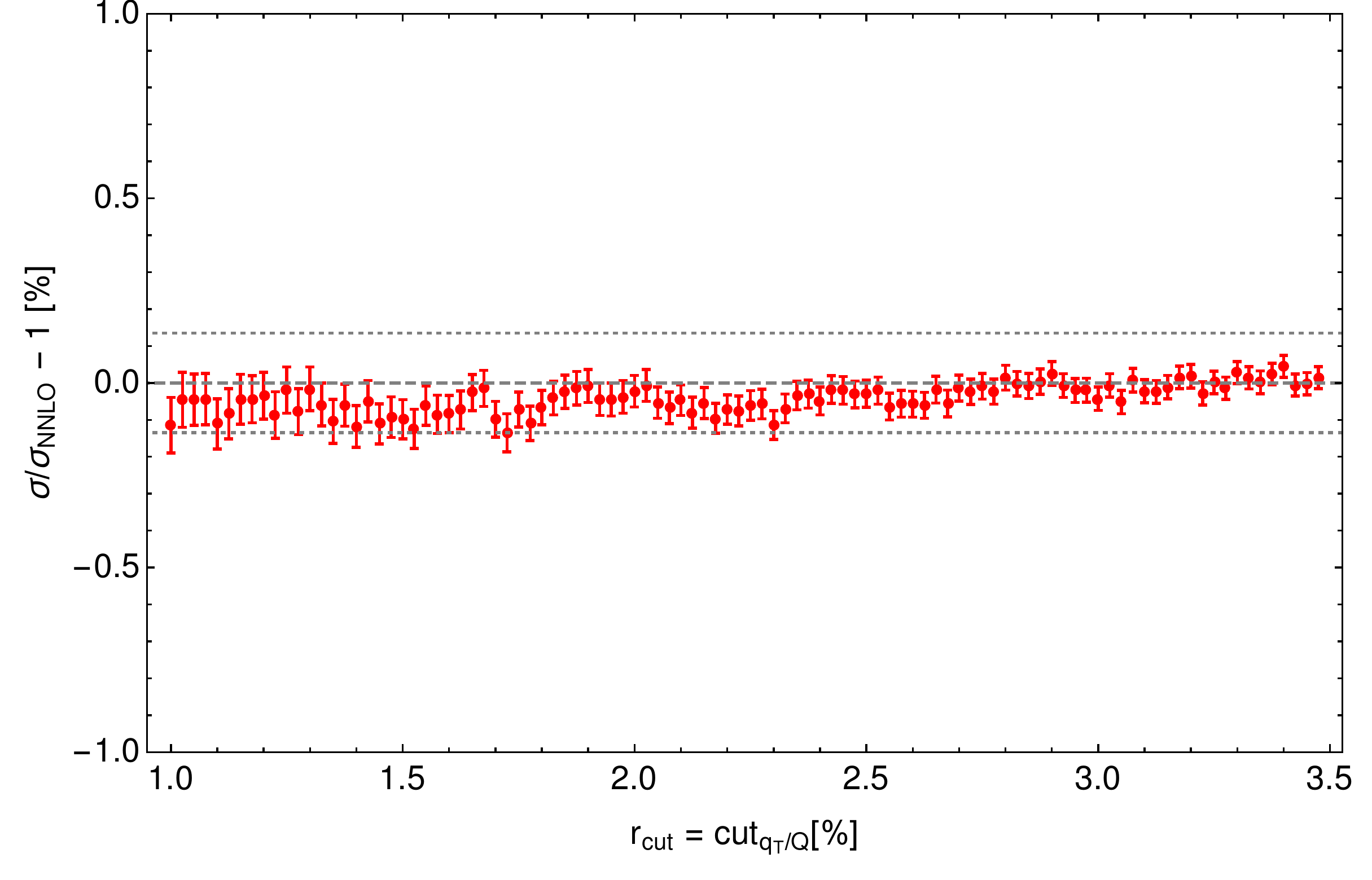}
\end{center}
\vspace{-0.7cm}
\caption{\label{fig:qT_cut}\small
Dependence of the total NNLO$_\text{FTapprox}$ cross section at $14\,$TeV on the $q_T$-subtraction cut, $r_\text{cut}$, normalized with respect to the extrapolated $r_\text{cut}\to 0$ result. The dotted lines indicate the symmetrized uncertainty coming from the extrapolation.
}
\end{figure}
%
In \reffi{fig:qT_cut} we illustrate the resulting dependence of the NNLO$_\text{FTapprox}$ total cross section on the $q_T$-subtraction cut, $r_\text{cut}$, for $\sqrt{s}=14\,$TeV.
Due to the previously discussed stability challenges, we considered values of $r_\text{cut}$ between $1\%$ and $3.5\%$, which are larger than the ones typically used in previous $q_T$-subtraction calculations (compared for instance with the default values in the public {\sc Matrix} release~\cite{Grazzini:2017mhc}). Nevertheless our results present a good stability, with effects that are below $0.2\%$ in the whole $q_T/Q$ range under study, validating this choice.
The $r_\text{cut}\to 0$ extrapolation is performed using a linear least $\chi^2$ fit. The fit is repeated varying the upper bound of the interval (in this case starting from a minimum of 25 points, which corresponds to an upper bound of $r_\text{cut} = 1.6\%$, and up to $r_\text{cut} = 3.5\%$).
Then, the result with the lowest $\chi^2/$degrees-of-freedom value is taken as the best fit, and the rest is used to estimate the extrapolation uncertainty.
In the case shown in \reffi{fig:qT_cut} the extrapolation uncertainty for $r_\text{cut}\to 0$, indicated with the dotted lines, is $\pm 0.14\%$.

A further uncertainty arises due to the numerical evaluation of the two-loop integrals with full top-quark mass dependence in the virtual corrections of the NLO contribution. 
The error of the numerical integration of the amplitudes is propagated to the total cross section using Monte Carlo methods, varying the amplitude level results according to the corresponding error estimates.
 This leads to changes of the NLO cross section at the per mille level. Furthermore, we have checked that, within this uncertainty, results based on the grid for the virtual amplitude are consistent with  the ones directly obtained from the amplitude results calculated in \citeres{Borowka:2016ehy,Borowka:2016ypz}.
We want to point out that the uncertainties can be somewhat larger in differential results, in particular in the tails of $p_T$ and invariant-mass distributions.

This discussion shows that the uncertainties due to the $q_T$-subtraction method and the numerical evaluation of the NLO virtual contribution and grid interpolation are clearly under sufficient control.

%% file: results.tex
In this section we present our numerical predictions for inclusive and differential cross sections for Higgs boson pair production in $pp$ collisions.
We consider centre-of-mass energies of $13$, $14$, $27$ and $100\,$TeV.
For the sake of brevity, differential distributions are presented only for $14\,$TeV and $100\,$TeV.
We use the values $M_h = 125\,$GeV and $M_t = 173\,$GeV for the Higgs boson and top quark masses, respectively. We do not consider bottom quark loops, whose contribution at LO is below $1\%$.
We also neglect top quark width effects, which at LO are at the level of $2\%$ for the total cross section~\cite{Maltoni:2014eza}.
We use the PDF4LHC15 sets~\cite{Butterworth:2015oua,Ball:2014uwa,Dulat:2015mca,Harland-Lang:2014zoa,Gao:2013bia,Carrazza:2015aoa}
of parton distribution functions (PDFs), with parton densities
and $\as$ evaluated at each corresponding perturbative order (i.e., we use the $(k + 1)$-loop running
$\as$ at N$^k$LO, with $k=1,2$).
As renormalization and factorization scales, we use the central value $\mu_0 = M_{hh}/2$, and we obtain scale uncertainties via the usual 7-point scale variation.

\subsection{Inclusive cross sections}
\label{sec:inclusive}

In \refta{table:totalXS} we present results for the total cross sections at NLO and NNLO in the various approximations.
At NLO we report the exact result, including the full $M_t$ dependence, and also the \ftapprox result.
By comparing the two NLO predictions, we see that the FT approximation overestimates the exact NLO result by 4$\%$ (6$\%$) at $14$ $(100)\,$TeV.
At NNLO the largest prediction is obtained in the NNLO$_\text{B-proj}$ approximation, resulting in an increase with respect to the exact NLO result of about $20\%$ at $14\,$TeV.
For this collider energy, the increase within the NNLO$_\text{NLO-i}$ approach (which is computed based on the $M_{hh}$ distribution) is smaller, being about $18\%$.
Finally, the NNLO$_\text{FTapprox}$ prediction is the lowest one, with a $12\%$ increase with respect to the NLO cross section at $14\,$TeV. 
For all the considered approximations and collider energies the scale uncertainties are significantly reduced when including the ${\cal O}(\as^4)$ NNLO corrections. This reduction is largest for the \nnloBP and \nnloFT approximations\footnote{The scale uncertainty of the NNLO$_\text{NLO-i}$ prediction is defined as the relative uncertainty of the HEFT result.}.
For instance at $14\,$TeV, the total scale uncertainty is reduced from about $\pm 13\%$ at NLO to $+2\%-5\%$ at \nnloFT, i.e. by about a factor of  three.
This reduction of the scale uncertainties is stronger as we increase the collider energy, being close to a factor of five at $100\,$TeV.

As is well known, scale uncertainties can only provide a lower limit on the true perturbative uncertainties. In particular, from \refta{table:totalXS} we see that
the difference between the NNLO and NLO central predictions is always larger than the NNLO scale uncertainties (although within the NLO uncertainty bands).
In any case, the strong reduction of scale uncertainties, together with the moderate impact of NNLO corrections, suggests a significant improvement in the perturbative convergence as we move from NLO to NNLO.
%

{\renewcommand{\arraystretch}{1.6}
\begin{table}[t]
\begin{center}
\begin{tabular}{l|c|c|c|c}
$\sqrtS$  & 13 TeV & 14 TeV  & 27 TeV & 100 TeV
\\[0.5ex]
\hline
\hline
\Tstrut
NLO [fb]           
& $27.78\,^{+13.8\%}_{-12.8\%}$ 
& $32.88\,^{+13.5\%}_{-12.5\%}$ 
& $127.7\,^{+11.5\%}_{-10.4\%}$ 
& $1147\,^{+10.7\%}_{-9.9\%}$ 
\Bstrut\\
\Tstrut
NLO$_\text{FTapprox}$ [fb]           
& $28.91\,^{+15.0\%}_{-13.4\%}$ 
& $34.25\,^{+14.7\%}_{-13.2\%}$ 
& $134.1\,^{+12.7\%}_{-11.1\%}$ 
& $1220\,^{+11.9\%}_{-10.6\%}$ 
\Bstrut\\
\hline
\Tstrut
\nnloNI [fb]           
& $32.69\,^{+5.3\%}_{-7.7\%}$ 
& $38.66\,^{+5.3\%}_{-7.7\%}$ 
& $149.3\,^{+4.8\%}_{-6.7\%}$ 
& $1337\,^{+4.1\%}_{-5.4\%}$ 
\Bstrut\\
\Tstrut
\nnloBP  [fb]           
& $33.42\,^{+1.5\%}_{-4.8\%}$ 
& $39.58\,^{+1.4\%}_{-4.7\%} $  
& $154.2\,^{+0.7\%}_{-3.8\%}$  
& $1406\,^{+0.5\%}_{-2.8\%}$  
\Bstrut\\
\Tstrut
\nnloFT [fb]           
& $31.05\,^{+2.2\%}_{-5.0\%} $
& $36.69\,^{+2.1\%}_{-4.9\%} $ 
& $139.9\,^{+1.3\%}_{-3.9\%} $ 
& $1224\,^{+0.9\%}_{-3.2\%} $ 
\Bstrut\\
\hline
\Tstrut
$M_t$ unc. \nnloFT           
& $\pm 2.6\% $ 
& $\pm 2.7\% $  
& $\pm 3.4\% $ 
& $\pm 4.6\% $ 
\Bstrut\\
\hline
\Tstrut
\nnloFT/NLO         
& $1.118$
& $1.116$
& $1.096$
& $1.067$
\Bstrut\\
[0.5ex]\hline
\end{tabular}
\end{center}
\caption{
Inclusive cross sections for Higgs boson pair production for different centre-of-mass energies at NLO and NNLO within the three considered approximations. 
Scale uncertainties are reported as superscript/subscript.
The estimated top quark mass uncertainty of the \nnloFT predictions is also presented.
The uncertainties due to the $q_T$-subtraction and the numerical evaluation of the virtual NLO contribution are both at the per mille level.
}
\label{table:totalXS}
\end{table}

It is also worth mentioning that the three approximations have a different behaviour with $\sqrt{s}$. For instance at $100\,$TeV, the increase with respect to the NLO prediction for the NNLO$_\text{B-proj}$ and NNLO$_\text{NLO-i}$ approaches is $23\%$ and $17\%$, respectively, values that are close to the ones for $14\,$TeV ($20\%$ and $18\%$, respectively). By contrast, the NNLO$_\text{FTapprox}$ result increases the NLO prediction by $7\%$ at $100\,$TeV, i.e. the correction is smaller by almost a factor of two than at $14\,$TeV~($12\%$), which also means a larger separation with respect to the other two NNLO approximations.
The smaller size of the NNLO corrections in the \ftapprox at higher energies is also consistent with the observed reduction of scale uncertainties.

As was mentioned already in \refse{sec:approximations}, the \nnloFT result is expected to be the most accurate one among the approximations studied in this work, and therefore it is considered to be our best prediction.
In order to estimate the remaining uncertainty associated with finite top quark mass effects at NNLO, we start by considering
 the accuracy of the \ftapprox approximation at NLO.
At $14\,$TeV the NLO \ftapprox result (see \refta{table:totalXS}) overestimates the full NLO total cross section by only about $4\%$, or equivalently by about $11\%$ of the pure ${\cal O}(\as^3)$ contribution.
If we assume that \ftapprox performs analogously at one order higher, we obtain a $\pm 11\%$ uncertainty on the ${\cal O}(\as^4)$ contribution\footnote{
We point out that in order to obtain the pure ${\cal O}(\as^4)$ corrections, we have subtracted the lower order contributions computed with NNLO parton distributions and strong coupling. The corresponding numbers are a few percent lower than the ones given in \refta{table:totalXS} for the NLO results.}.
Given that the relative weight of the ${\cal O}(\as^4)$ contributions to the total NNLO cross section is definitely smaller than the weight of the ${\cal O}(\as^3)$ contributions to the NLO cross section, we obtain a significantly smaller overall uncertainty, in this case of $\pm 1.2\%$.
In order to be conservative, we can increase this estimate by a factor of two.
The relative difference between the \ftapprox and the full NLO result slightly increases with the collider energy.
However, at the same time the relative size of the ${\cal O}(\as^4)$ correction decreases. 
The NNLO uncertainty obtained with this procedure ranges from $\pm 2.3\%$ at $13\,$TeV to $\pm 3.1\%$ at $100\,$TeV. 

We can repeat the above procedure to estimate the uncertainty of the \nnloBP approximation,  which displays the largest differences with respect to the NNLO$_\text{FTapprox}$ result.
Similarly to what we do for \ftapprox, we can assign an uncertainty to the \nnloBP result by relying on the accuracy of the same approximation at NLO,
and conservatively multiplying by a factor of two.
The ensuing uncertainties range from $\pm 14\%$ at $\sqrt{s}=13\,$TeV to $\pm 36\%$ at $\sqrt{s}=100\,$TeV.
We find that the \nnloFT  prediction (always evaluated at $\mu_R = \mu_F = \mu_0$) is fully contained in the \nnloBP uncertainty band.
Actually, there is a large overlap between the two approximations, which includes in all the cases the central value of the \nnloFT, even when the conservative factor of two is not included.
This can be regarded as a non-trivial consistency check for our procedure.
We may be tempted to conclude our discussion by adopting the above procedure for the uncertainty estimate of our \nnloFT result.

However, we have already pointed out  that, as $\sqrt{s}$ increases, the difference between the \nnloFT and the other approximations increases. In particular, the difference
between the \nnloFT result and our ``next-to-best'' NNLO prediction, \nnloNI, is $5.2\%$ at $\sqrt{s}=13\,$TeV, and it becomes $9.2\%$ at $\sqrt{s}=100\,$TeV.
The significant increase of this difference with the collider energy suggests us a more conservative approach. 
Our final estimate for the finite top quark mass uncertainty of our
\nnloFT result is defined as {\em half the difference between the \nnloFT and the \nnloNI approximations}, and
 is reported in \refta{table:totalXS} for the different values of
 $\sqrt{s}$. At $\sqrt{s}=13$ and $14\,$TeV these uncertainties are $\pm 2.6\%$ and $\pm 2.7\%$, and thus very similar
to the ones obtained with the method discussed above. At $\sqrt{s}=100\,$TeV, however, the uncertainty increases to $\pm 4.6\%$, which appears to be more conservative than the $\pm 3.1\%$ obtained with the previous procedure.

\subsection{Differential distributions}

In this section we present predictions for differential Higgs boson pair production at $14\,$TeV and $100\,$TeV. We consider the following kinematical distributions: the invariant mass ($M_{hh}$, \reffi{fig:mhh}) and rapidity ($y_{hh}$, \reffi{fig:yhh}) of the Higgs boson pair, the transverse momenta of the Higgs boson pair and the leading jet ($p_{{\rm T},hh}$ and $p_{{\rm T},j1}$, \reffitwo{fig:pT_hh}{fig:pT_j1}), the transverse momenta of the harder and the softer Higgs boson ($p_{{\rm T},h1}$ and $p_{{\rm T},h2}$, \reffitwo{fig:pT_h1}{fig:pT_h2}), and the azimuthal separation between the two Higgs bosons ($\Delta\phi_{hh}$, \reffi{fig:phi_hh}).
For the sake of clarity, we only show the scale uncertainty bands corresponding to the NLO and \nnloFT predictions.

\begin{figure}[t!]
\begin{center}
\includegraphics[width=.49\textwidth]{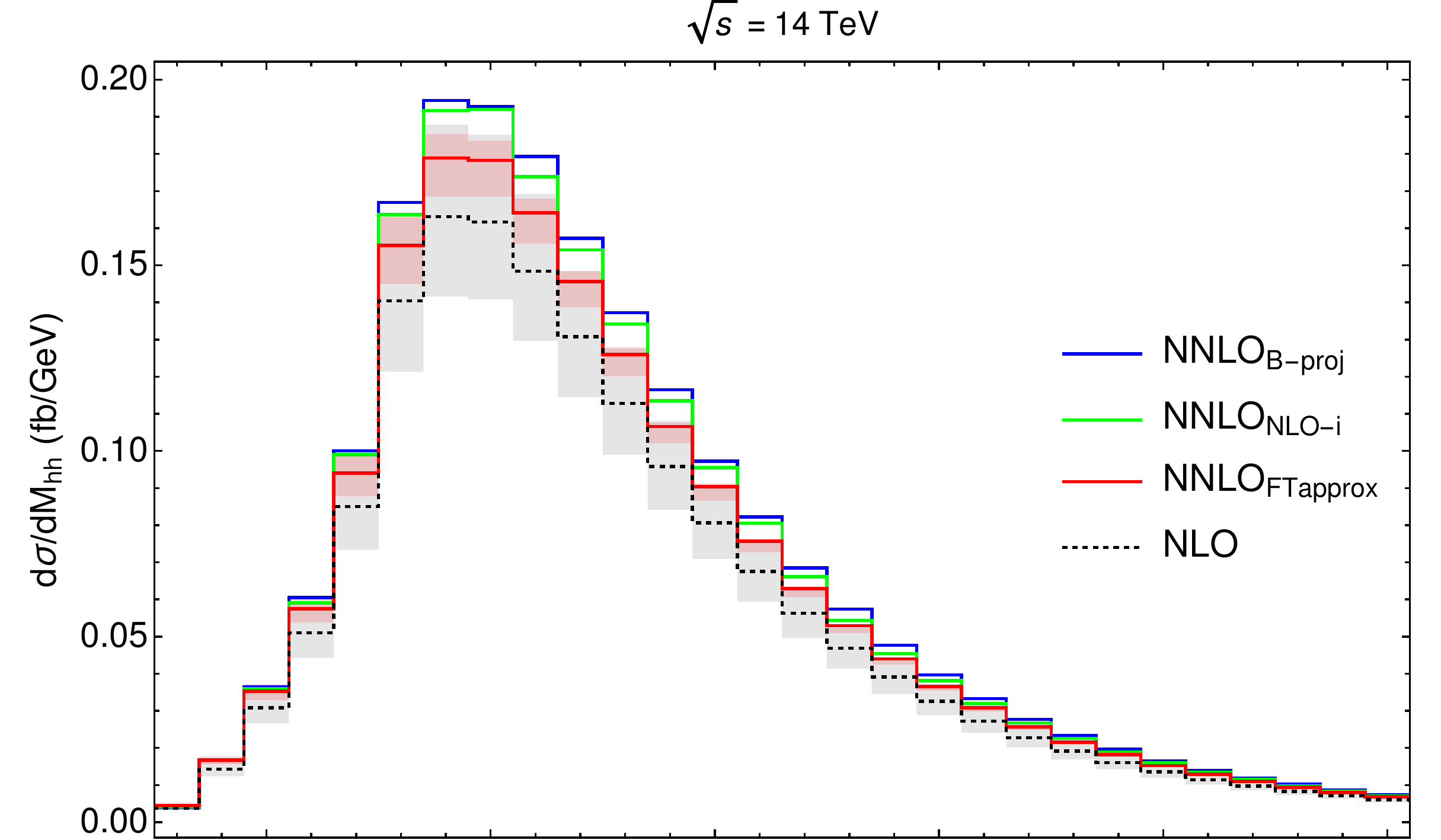}
\hfill
\includegraphics[width=.49\textwidth]{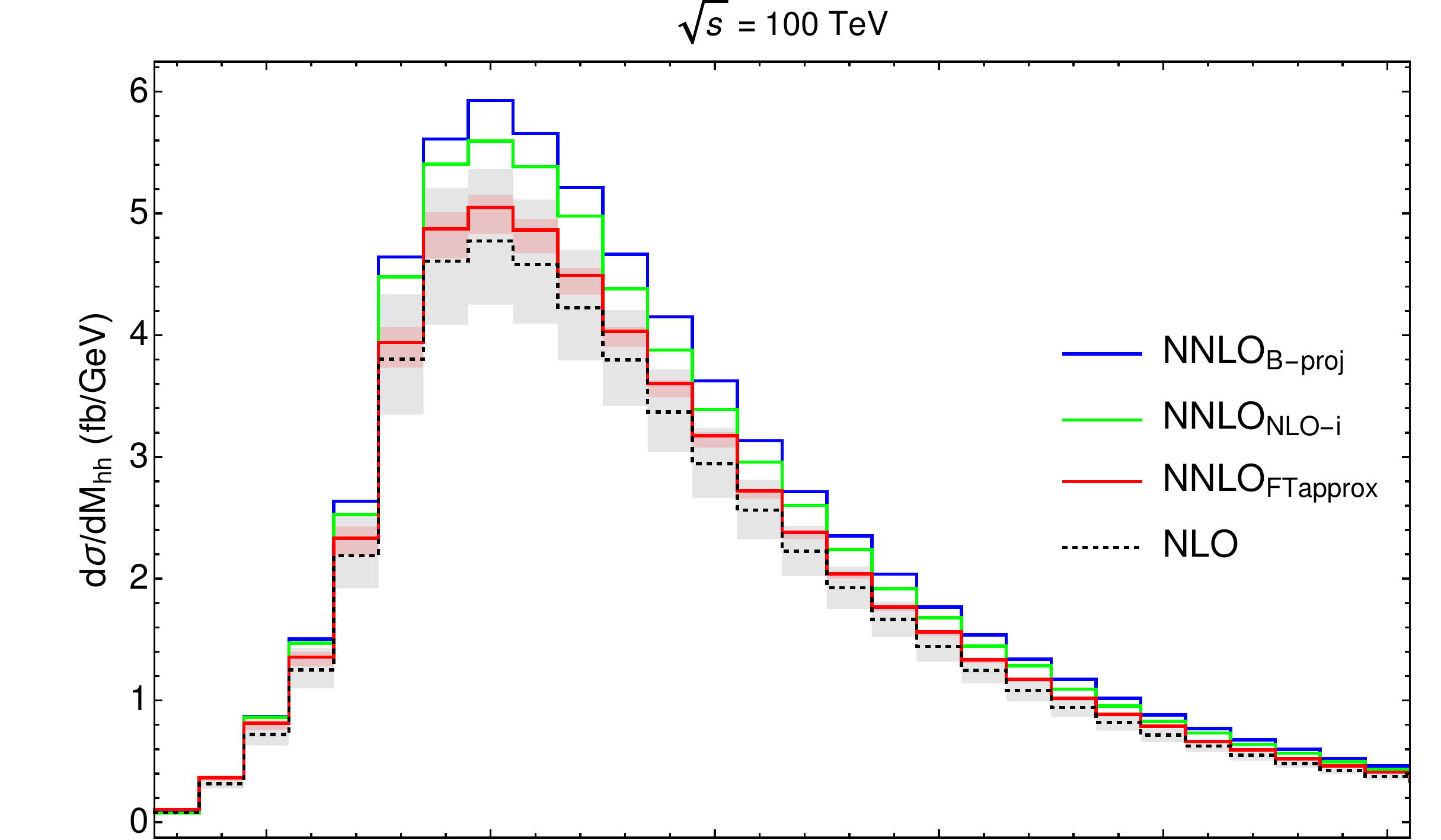}
\\
\includegraphics[width=.49\textwidth]{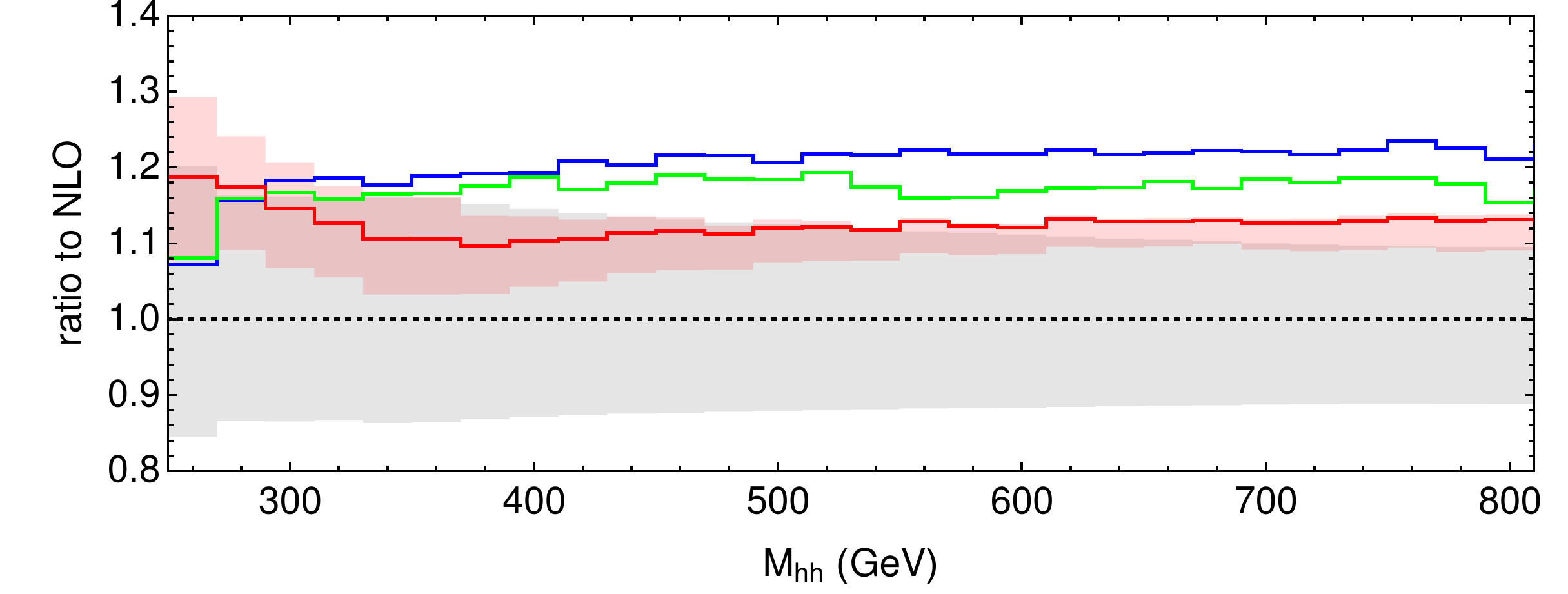}
\hfill
\includegraphics[width=.49\textwidth]{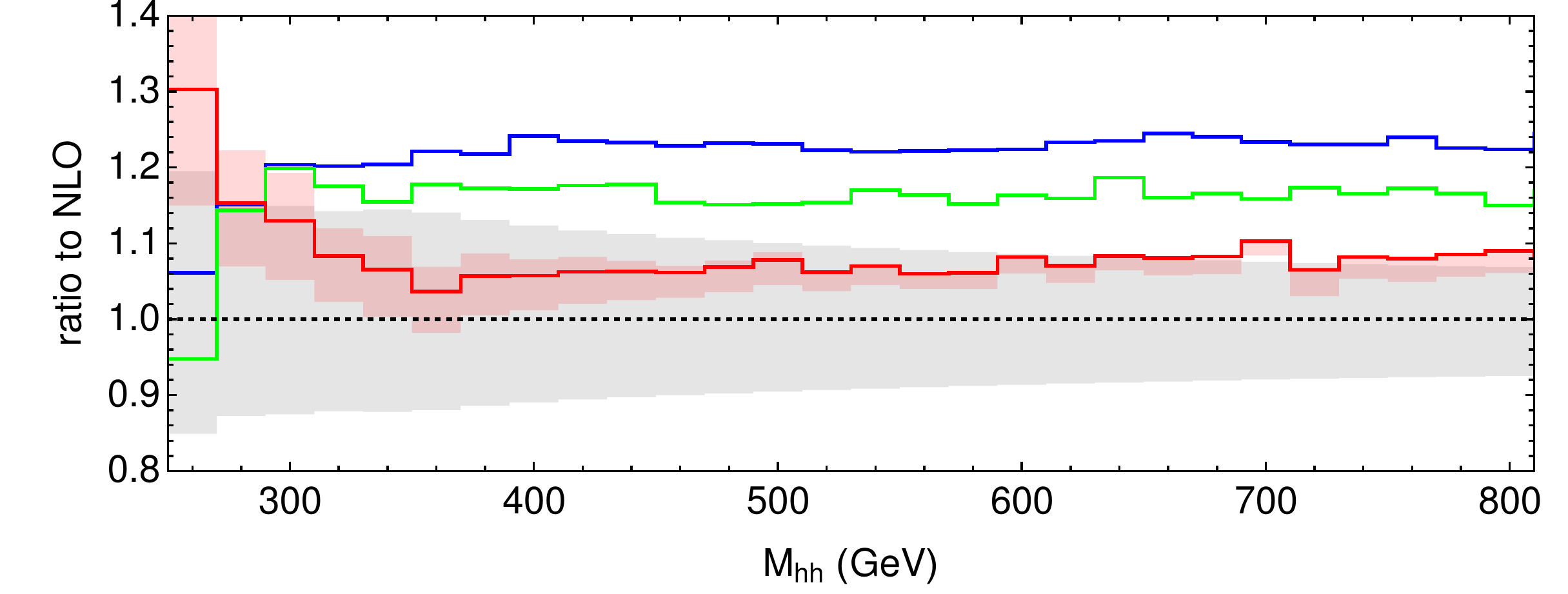}
\end{center}
\vspace{-2ex}
\caption{\label{fig:mhh}\small
Higgs boson pair invariant mass distribution at NNLO for the different approximations, together with the NLO prediction, at $14\,$TeV (left) and $100\,$TeV (right). The lower panels show the ratio with respect to the NLO prediction, and the filled areas indicate the NLO and NNLO$_\text{FTapprox}$ scale uncertainties.
}
\end{figure}

We start our discussion from the
invariant-mass distribution of the Higgs boson pair, reported in \reffi{fig:mhh}. We observe that the NNLO$_\text{B-proj}$ and NNLO$_\text{NLO-i}$ approximations predict a similar shape, with very small corrections at threshold, an approximately constant $K$-factor for larger invariant masses, and only a small difference in the normalization between them, which increases in the $100\,$TeV case.
The \nnloFT, on the other hand, presents a different shape, in particular with larger corrections for lower invariant masses, a minimum in the size of the corrections close to the region where the maximum of the distribution is located, and a slow increase towards the tail.
The different behavior of the \nnloFT in the region close to threshold is more evident at $100\,$TeV, where the increase is about 30\% in the first bin.
Naively we could expect that if this region is dominated by soft parton(s) recoiling against the Higgs bosons, the Born projection and \ftapprox should provide
similar results. We have investigated the origin of this difference, and we find that in the region $M_{hh}\sim 2M_h$ the cross section is actually dominated by events with relatively hard
radiation recoiling against the Higgs boson pair (for example, at $\sqrt{s}=100\,$TeV, the average transverse momentum of the Higgs boson pair in the first $M_{hh}$ bin is $p_{{\rm T},hh}\sim 100\,$GeV at NLO).
In this region the exact loop amplitudes behave rather differently as compared to the amplitudes evaluated in the HEFT: 
As the production threshold is approached, they go to zero faster than in the mass-dependent case, 
thus explaining the differences we find.
Within the \nnloFT, the corrections to the $M_{hh}$ spectrum range between $10\%$ and $20\%$ at $14\,$TeV.
The scale uncertainty is substantially reduced in the \nnloFT, and this reduction is particularly strong for large invariant masses. 
As observed at the inclusive level, the \nnloFT corrections are smaller at $100\,$TeV (except only for the first bin) and the difference with respect to the other approximations is larger.

\begin{figure}[t!]
\begin{center}
\includegraphics[width=.48\textwidth]{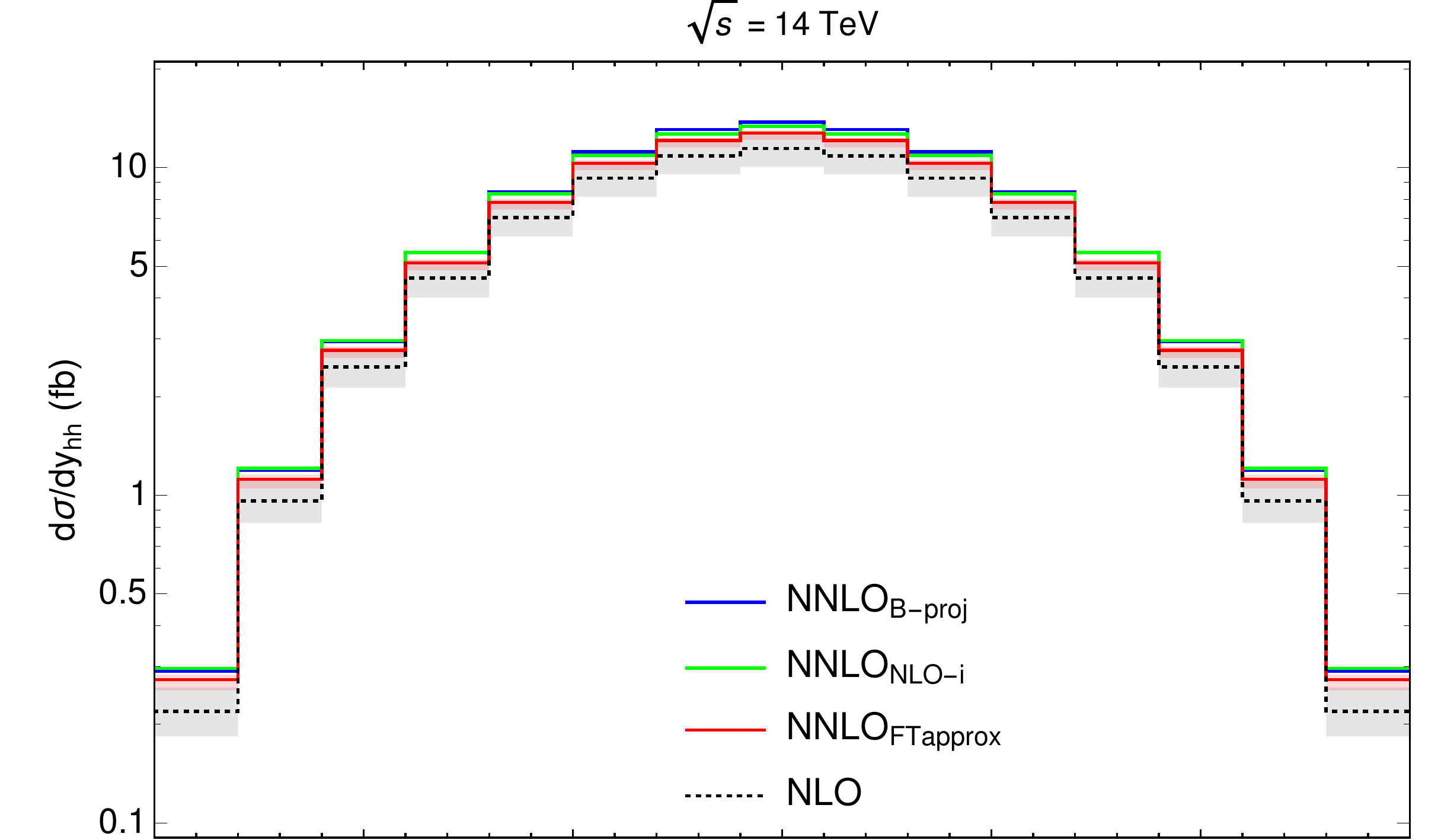}
\hfill
\includegraphics[width=.48\textwidth]{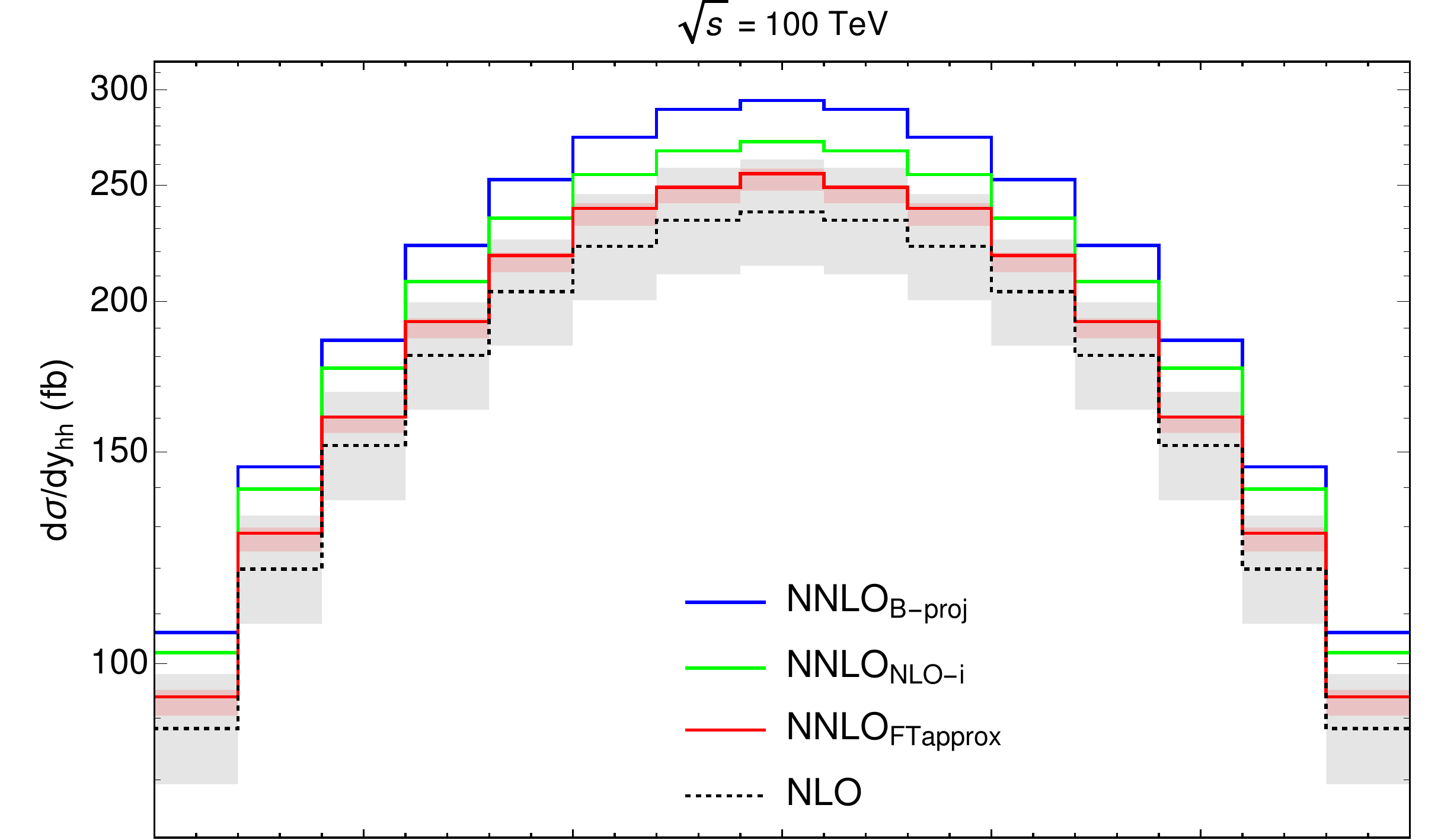}
\\
\includegraphics[width=.48\textwidth]{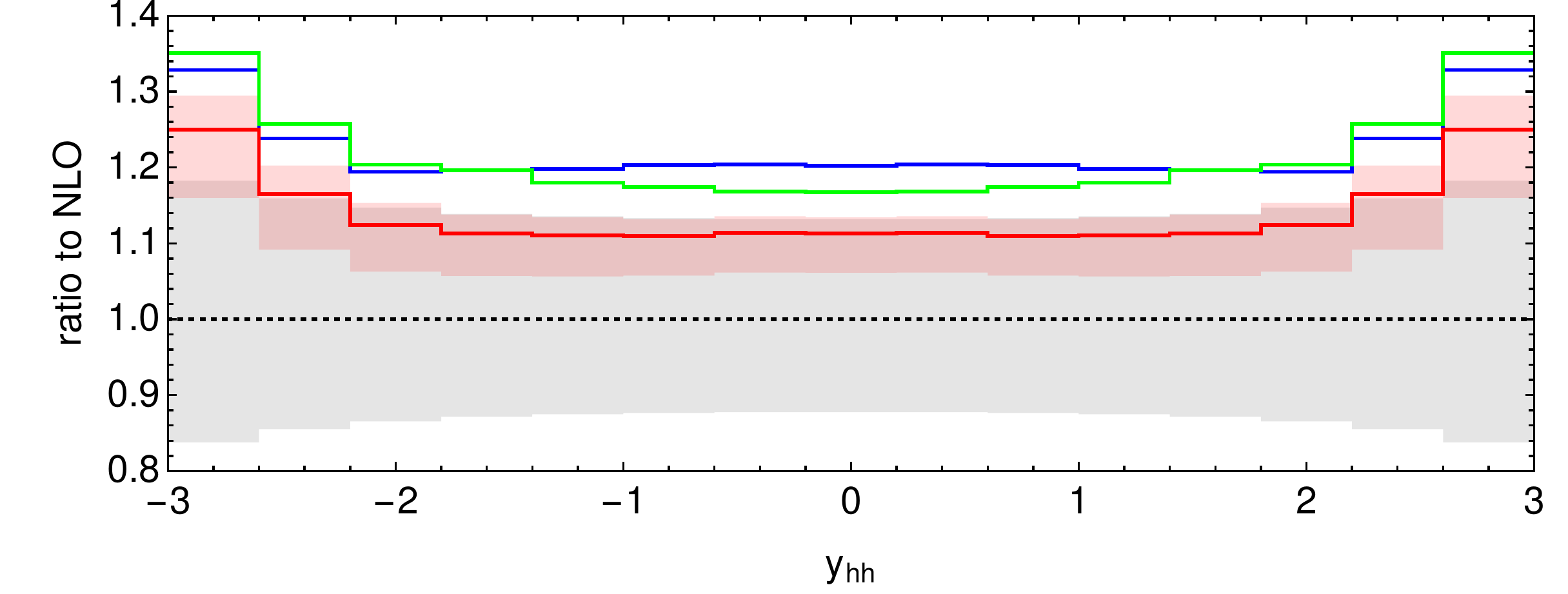}
\hfill
\includegraphics[width=.48\textwidth]{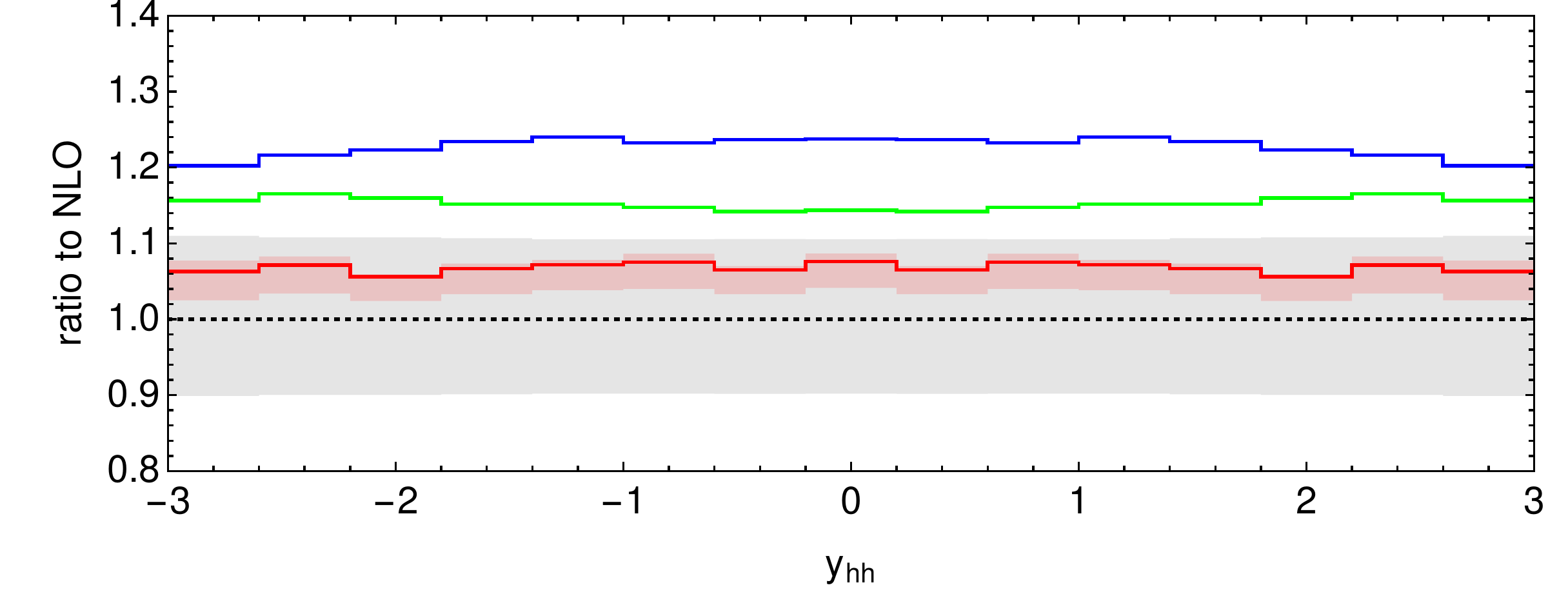}
\end{center}
\vspace{-2ex}
\caption{\label{fig:yhh}\small
Higgs boson pair rapidity distribution at NNLO for the different approximations, together with the NLO prediction, at $14\,$TeV (left) and $100\,$TeV (right).
}
\end{figure}

Next we move to the rapidity distribution of the Higgs boson pair, reported in \reffi{fig:yhh}. The NNLO results are similar for all three approximations. This is not unexpected as the shape of the rapidity distribution is mainly driven by the PDFs.
Besides the obvious difference in the normalization, the largest effect in the shape
of the \nnloNI distribution is observed in the central region, which is particularly evident in the $100\,$TeV case.
Again we observe a clear reduction of scale uncertainties over the whole range under study.

\begin{figure}[t!]
\begin{center}
\includegraphics[width=.48\textwidth]{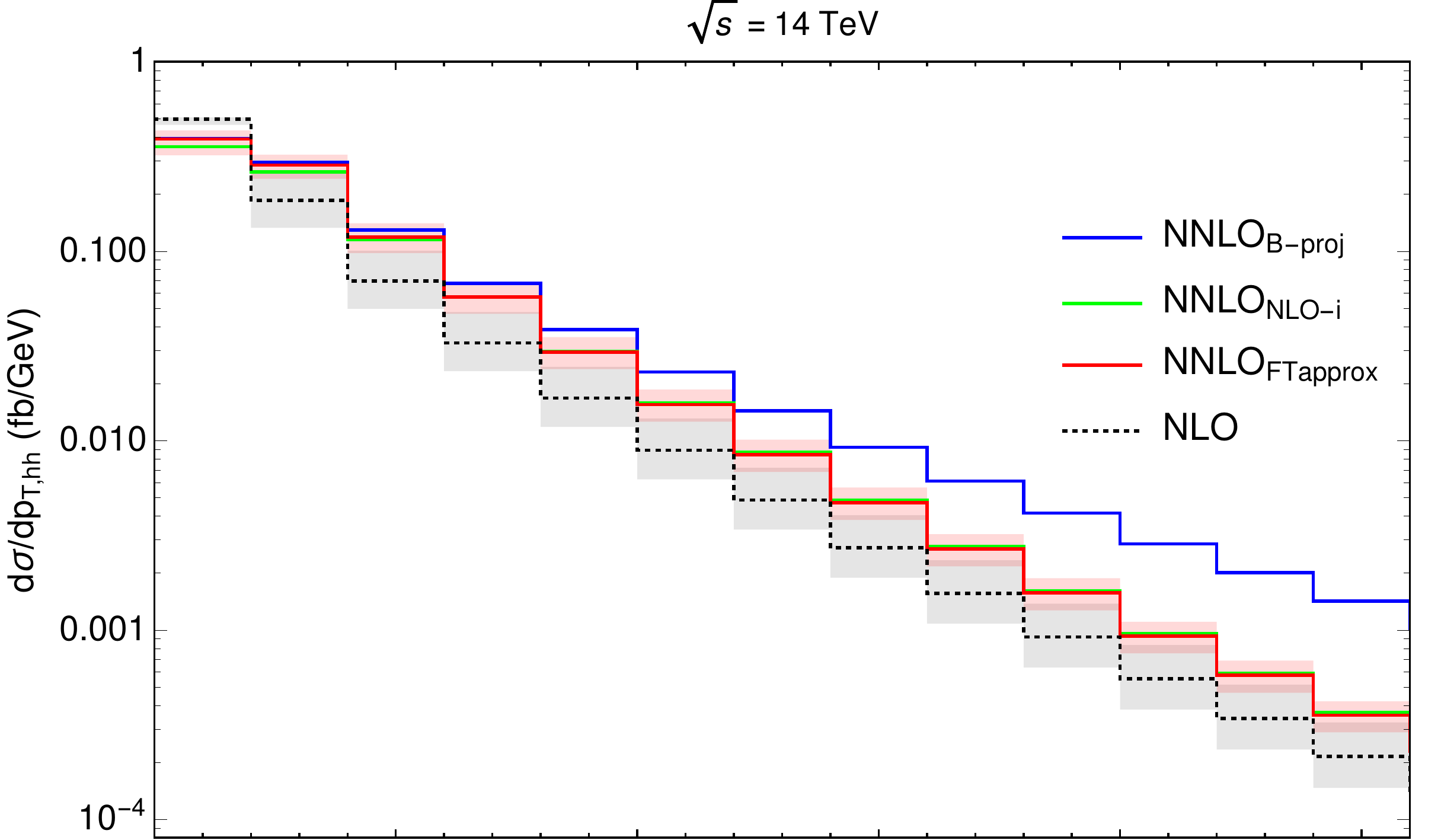}
\hfill
\includegraphics[width=.48\textwidth]{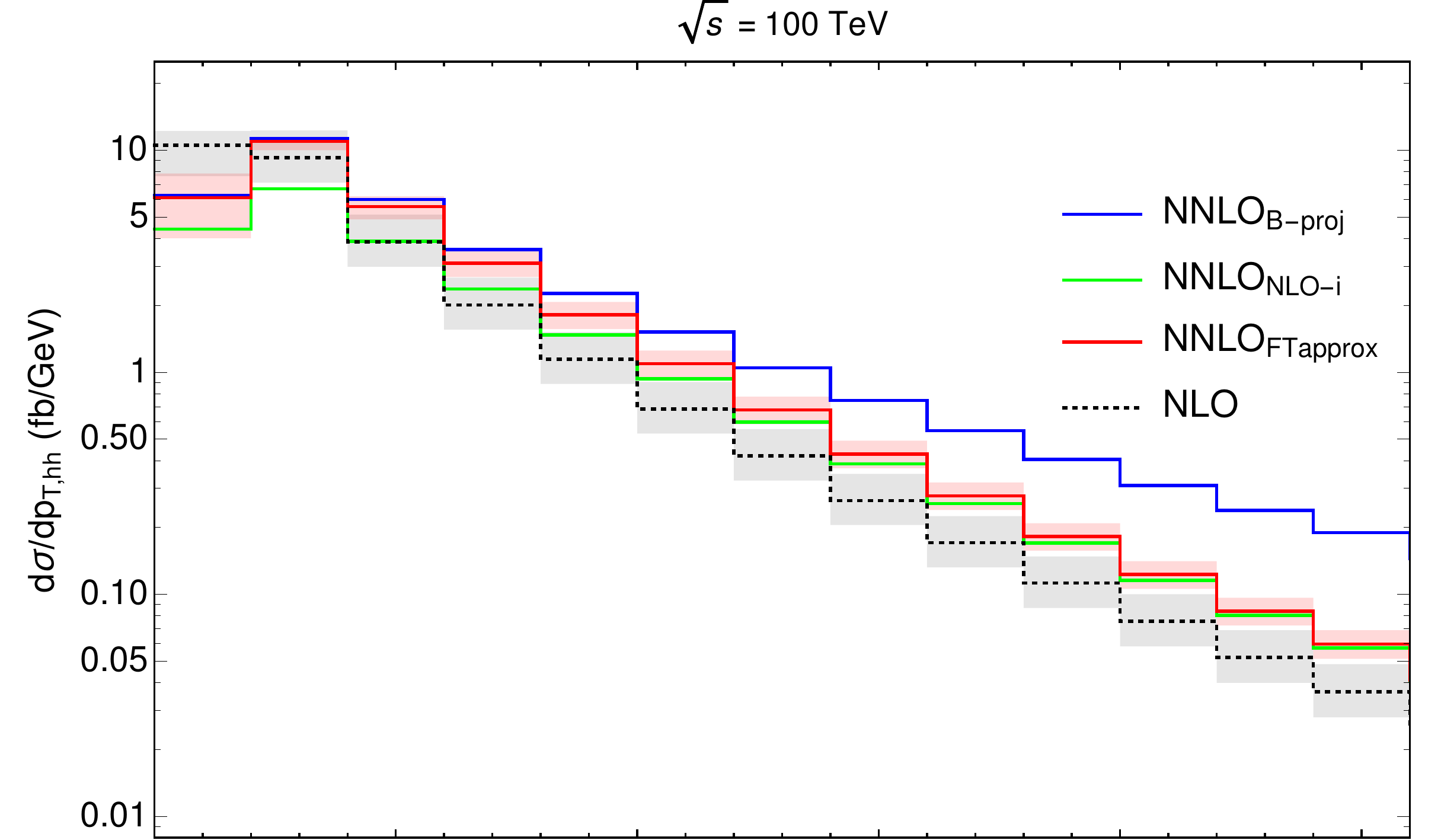}
\\
\includegraphics[width=.48\textwidth]{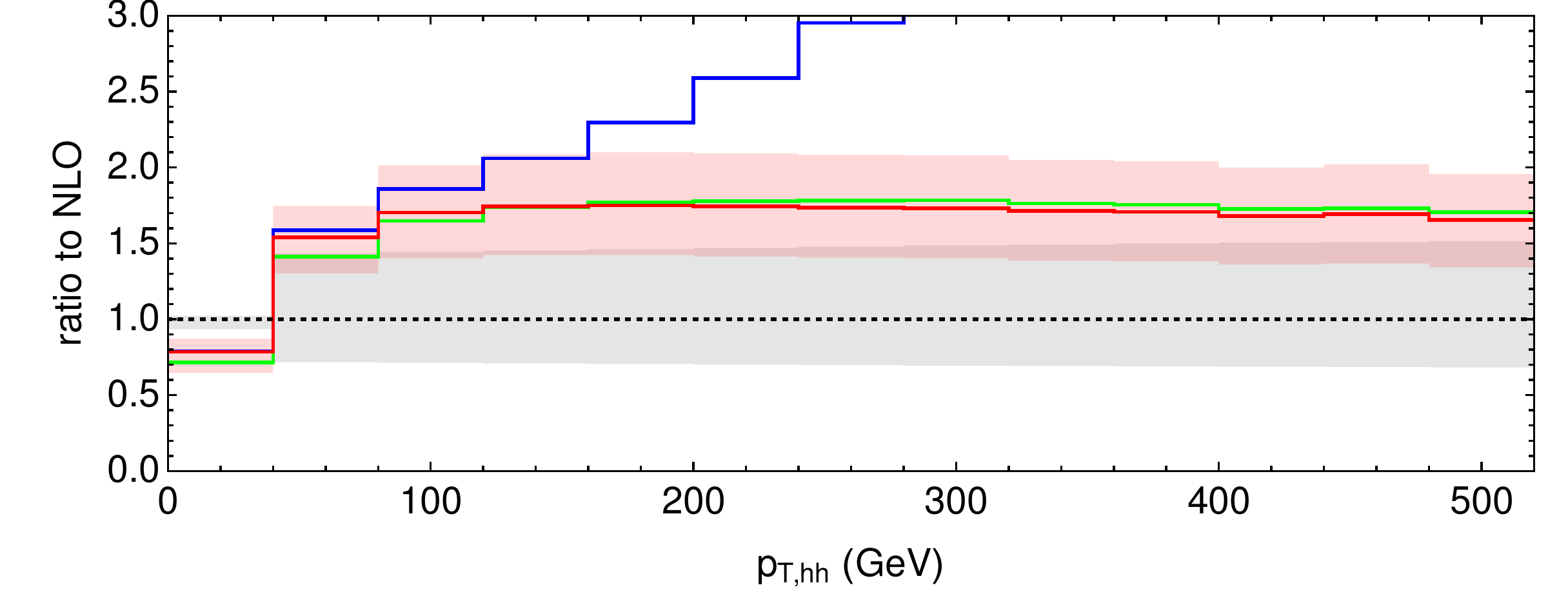}
\hfill
\includegraphics[width=.48\textwidth]{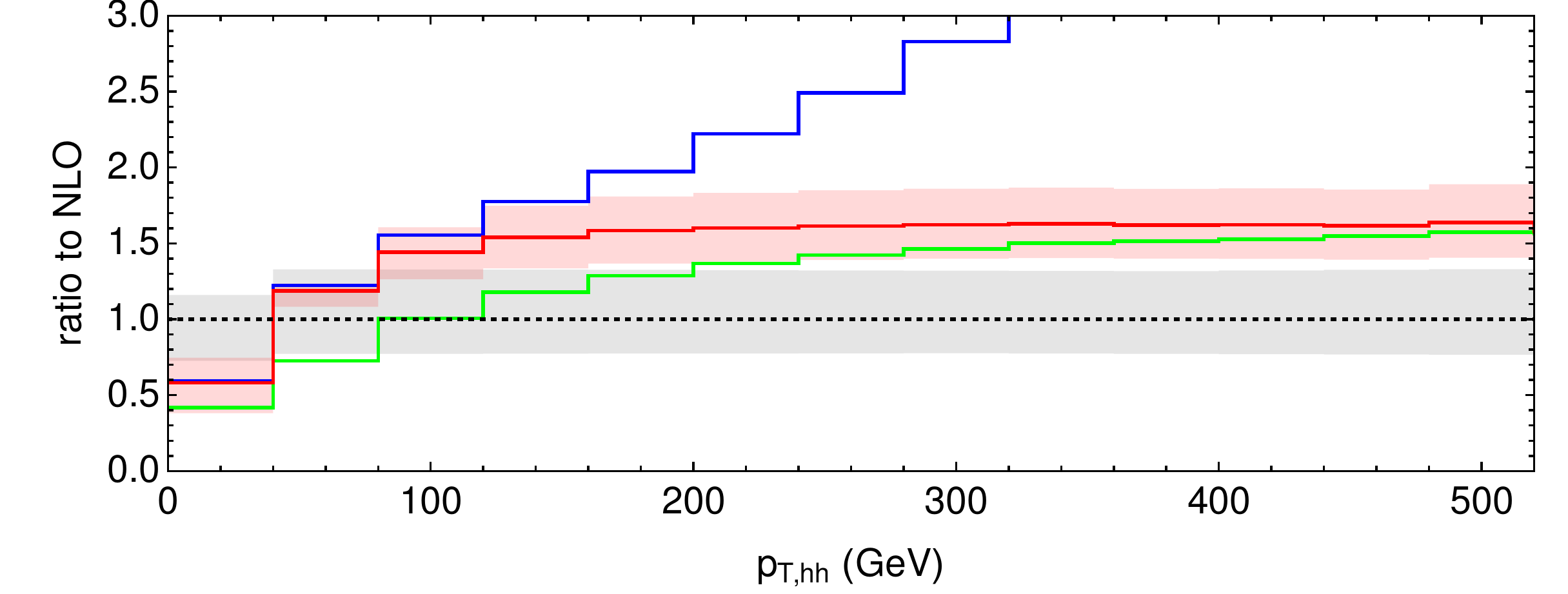}
\end{center}
\vspace{-2ex}
\caption{\label{fig:pT_hh}\small
Higgs boson pair transverse momentum distribution at $14\,$TeV (left) and $100\,$TeV (right).
}
\end{figure}

More significant differences between the three approximations are obtained in the $p_{{\rm T},hh}$ distribution, reported in \reffi{fig:pT_hh}. The NNLO$_\text{B-proj}$ approximation predicts huge corrections for large transverse momentum, the result being almost an order of magnitude larger than the NLO prediction and the other approximations for $p_{{\rm T},hh}\sim 500\,$GeV.
This behavior is hardly surprising since already at NLO the Born-projected result deviates from the exact NLO prediction in this way~\cite{Borowka:2016ypz}.
In fact, given that the $p_{{\rm T},hh}$ distribution is not defined at LO, the NNLO$_\text{B-proj}$ corrections cannot inherit any information about the (full) lowest-order prediction for this distribution.
This is of course not the case for the other two approximations, which in fact make an almost identical prediction at large $p_{{\rm T},hh}$, with large corrections that can be well above $50\%$, and sizable uncertainties at the level of $30\%$--$40\%$, reflecting the NLO-nature of this observable.
At lower transverse momenta, however, the \nnloNI and \nnloFT deviate from each other, and the latter approaches the \nnloBP prediction.
Once again, the different behavior of these approximations is more pronounced in the $100\,$TeV distribution, for which the central \nnloNI curve lies outside the \nnloFT uncertainty band below $p_{{\rm T},hh} \sim 200\,$GeV.
Of course, in order to obtain reliable results in the low-$p_{{\rm T},hh}$ region,
the corresponding logarithmically enhanced contributions need to be properly resummed to all orders in the strong coupling constant.

\begin{figure}[t!]
\begin{center}
\includegraphics[width=.48\textwidth]{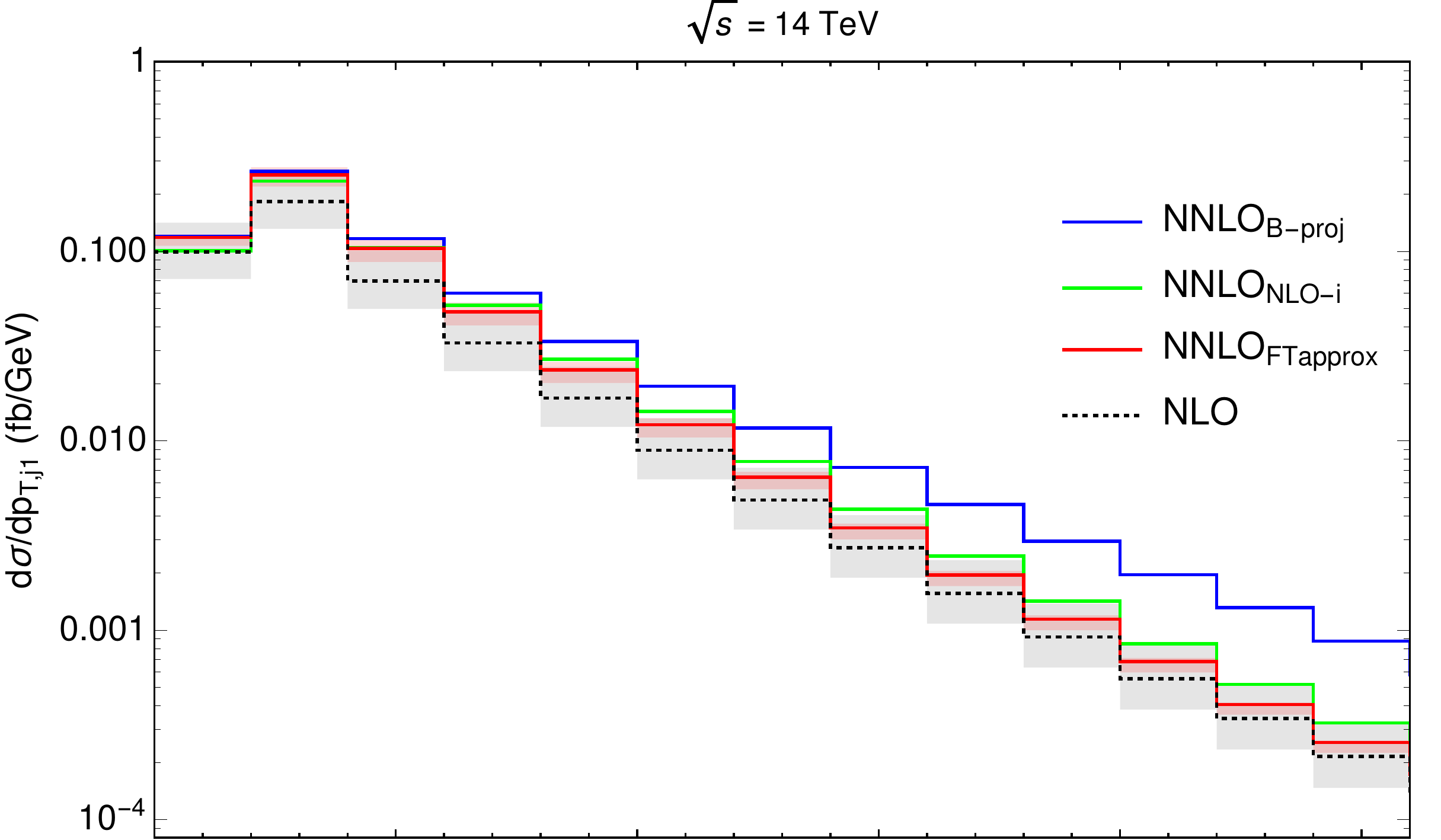}
\hfill
\includegraphics[width=.48\textwidth]{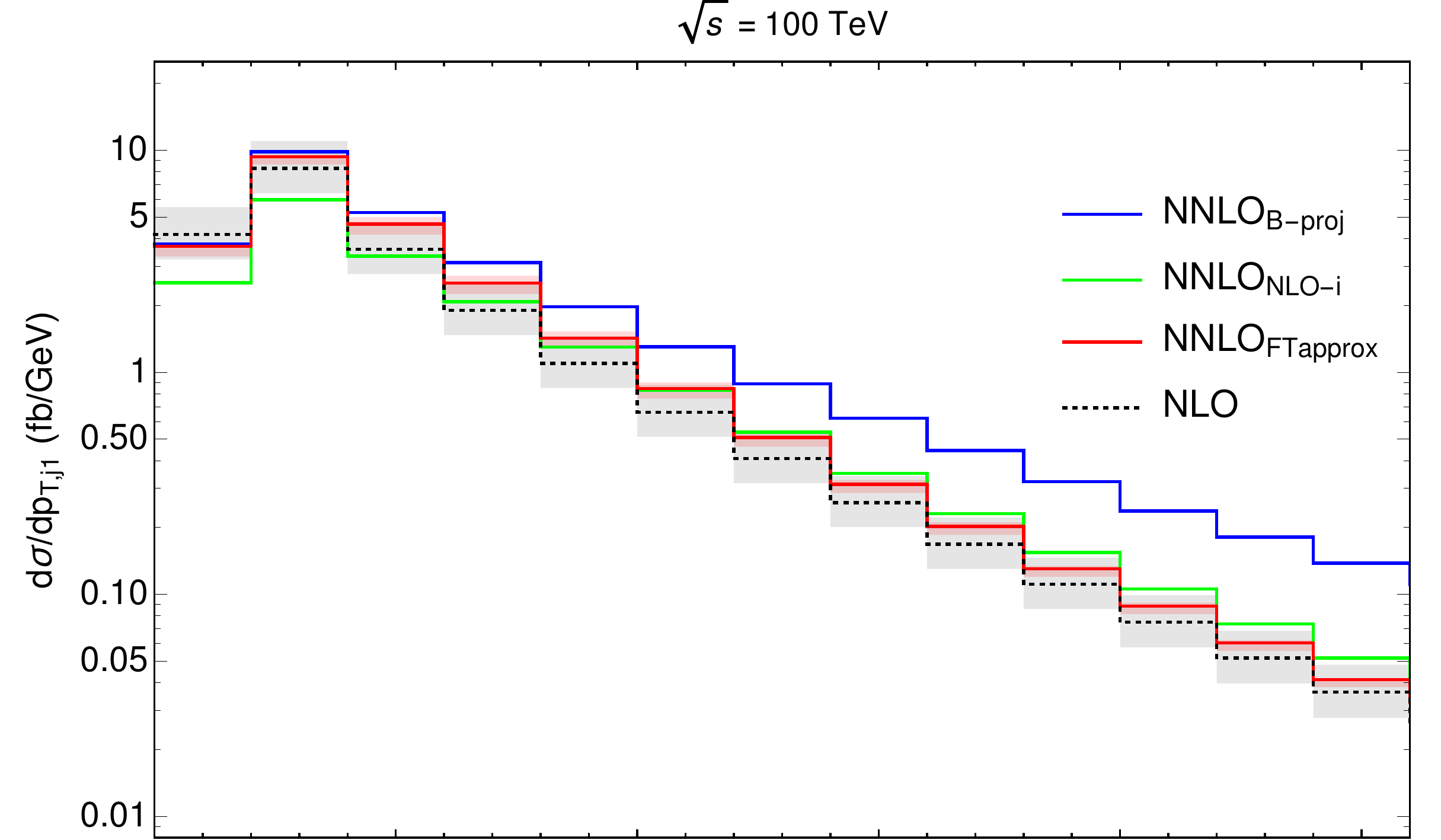}
\\
\includegraphics[width=.48\textwidth]{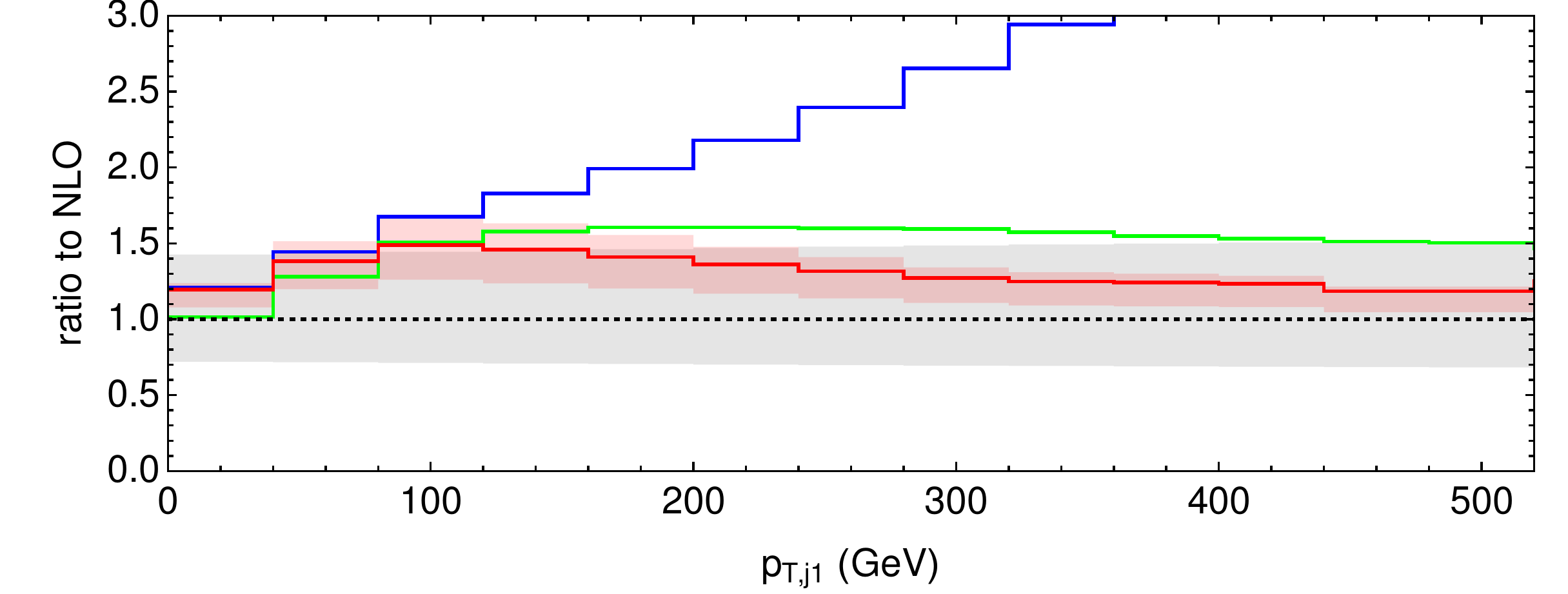}
\hfill
\includegraphics[width=.48\textwidth]{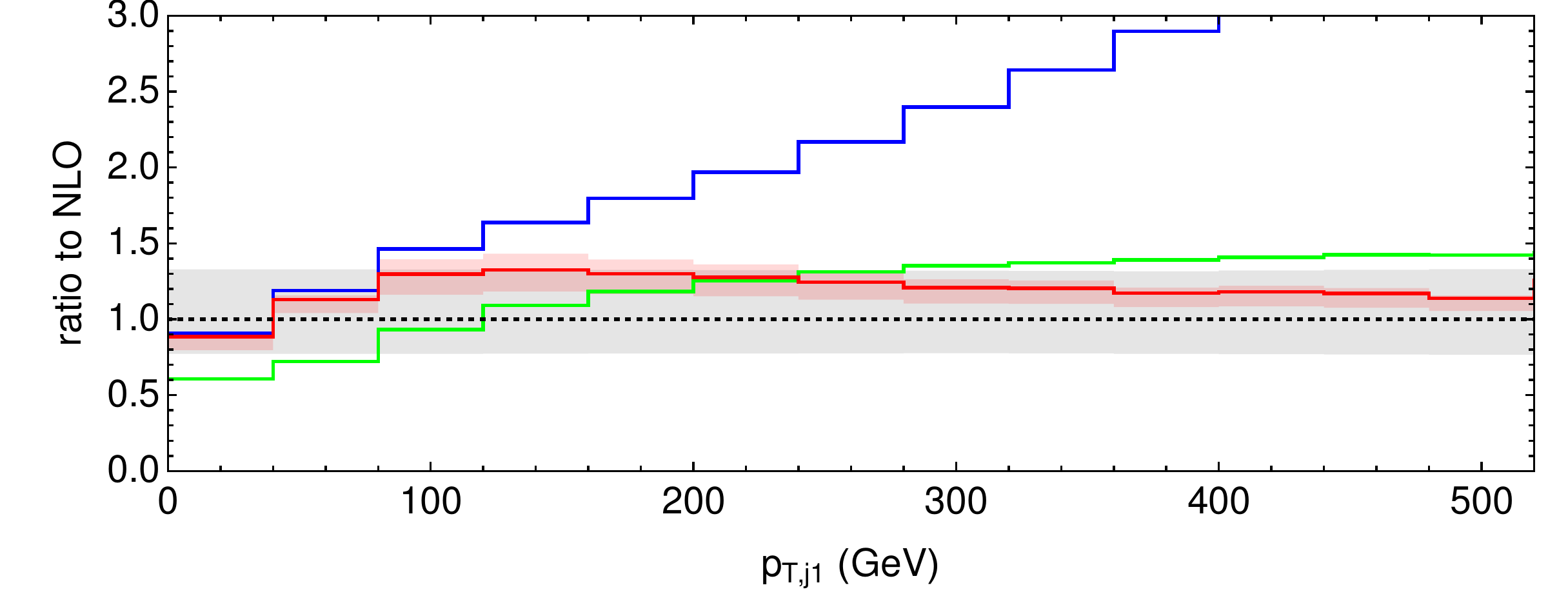}
\end{center}
\vspace{-2ex}
\caption{\label{fig:pT_j1}\small
Leading jet transverse momentum distributions at $14\,$TeV (left) and $100\,$TeV (right). 
Here jets are clustered with the anti-$k_T$ algorithm~\cite{Cacciari:2008gp} with $R=0.4$ and $p_{\rm{T},j1}>30\,$GeV and $|\eta_j|\leq 4.4$. 
}
\end{figure}

The transverse momentum distribution of the leading jet $p_{{\rm T},j1}$, reported in \reffi{fig:pT_j1},
%
%
has similar features as the $p_{{\rm T},hh}$ distribution. Again we observe the unphysical excess predicted by the \nnloBP approximation, which can be understood using the same arguments as presented for the $p_{{\rm T},hh}$ distribution, and the agreement between \nnloBP and \nnloFT at low $p_{{\rm T},j1}$.
The difference between the \nnloNI and \nnloFT results is more pronounced here, with the \ftapprox predicting a softer spectrum for this observable, and small corrections that are almost always contained in the NLO scale uncertainty band.

The transverse-momentum distributions of the harder and the softer Higgs boson are reported in \reffitwo{fig:pT_h1}{fig:pT_h2}, respectively.
\begin{figure}[t!]
\begin{center}
\includegraphics[width=.48\textwidth]{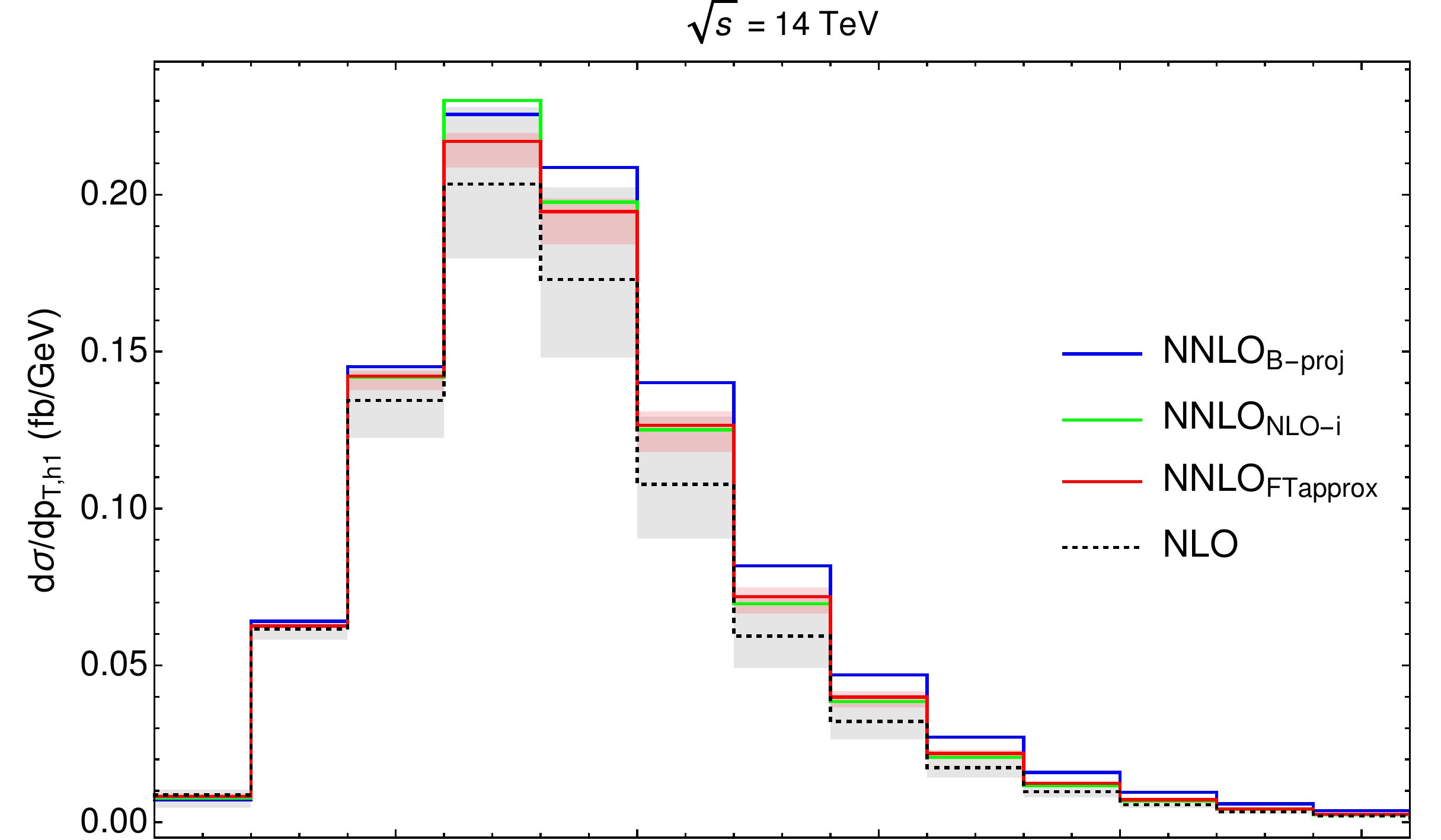}
\hfill
\includegraphics[width=.48\textwidth]{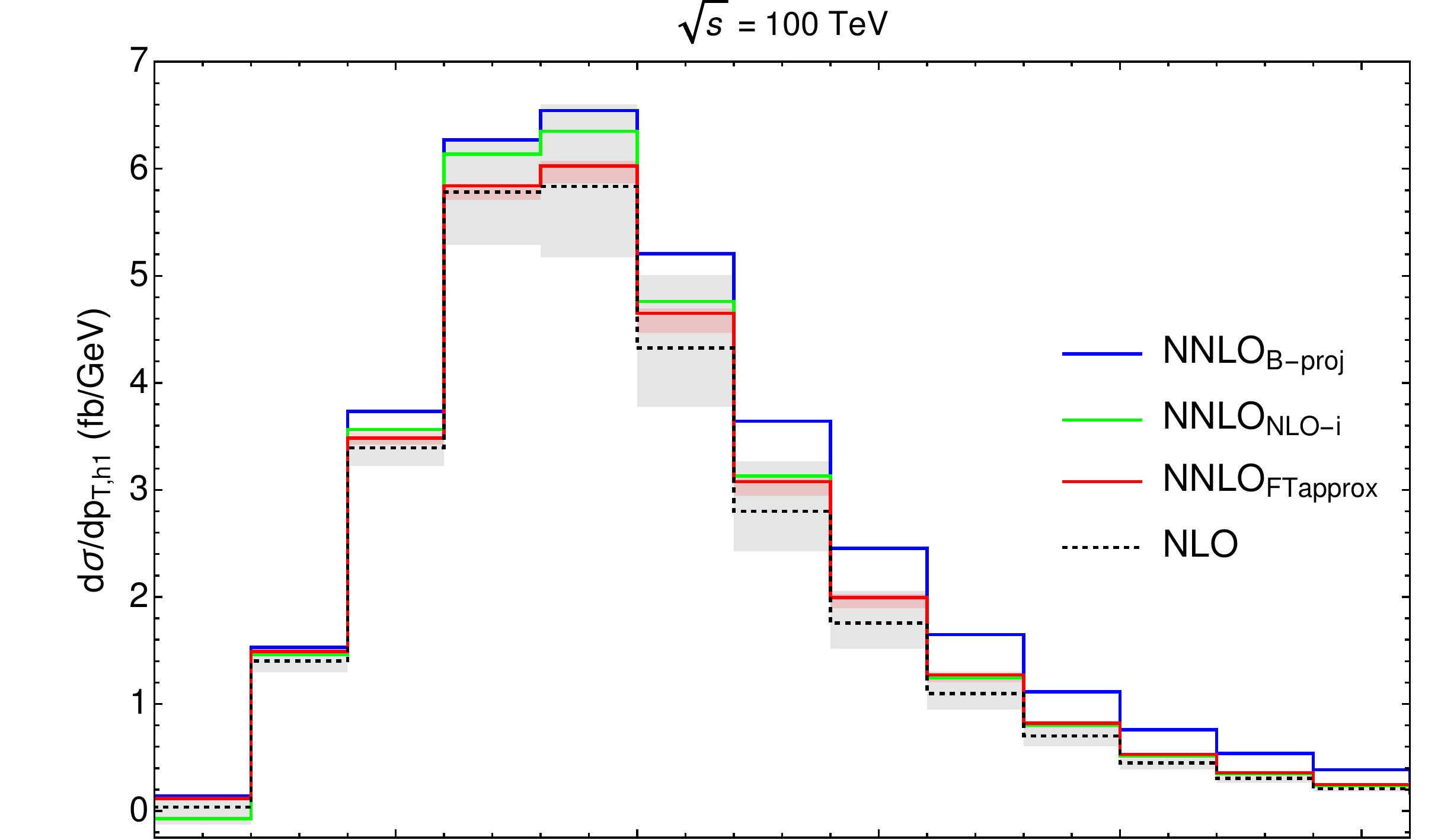}
\\
\includegraphics[width=.48\textwidth]{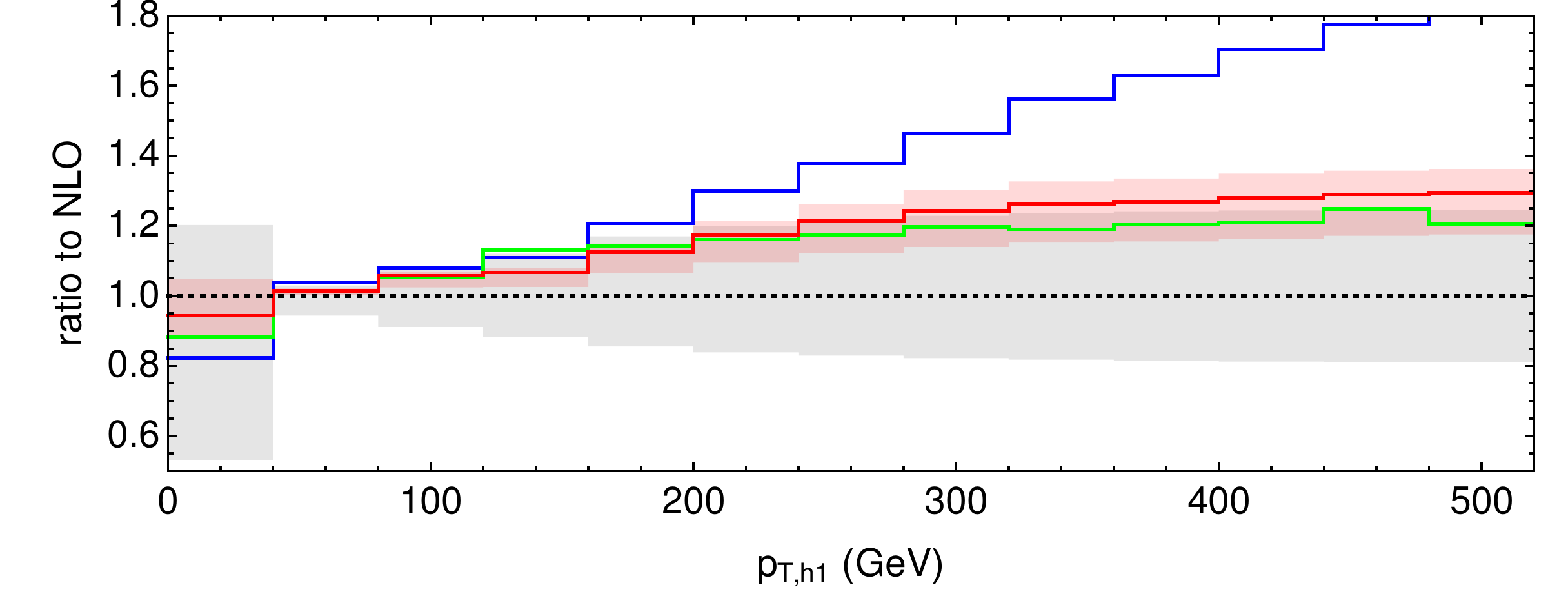}
\hfill
\includegraphics[width=.48\textwidth]{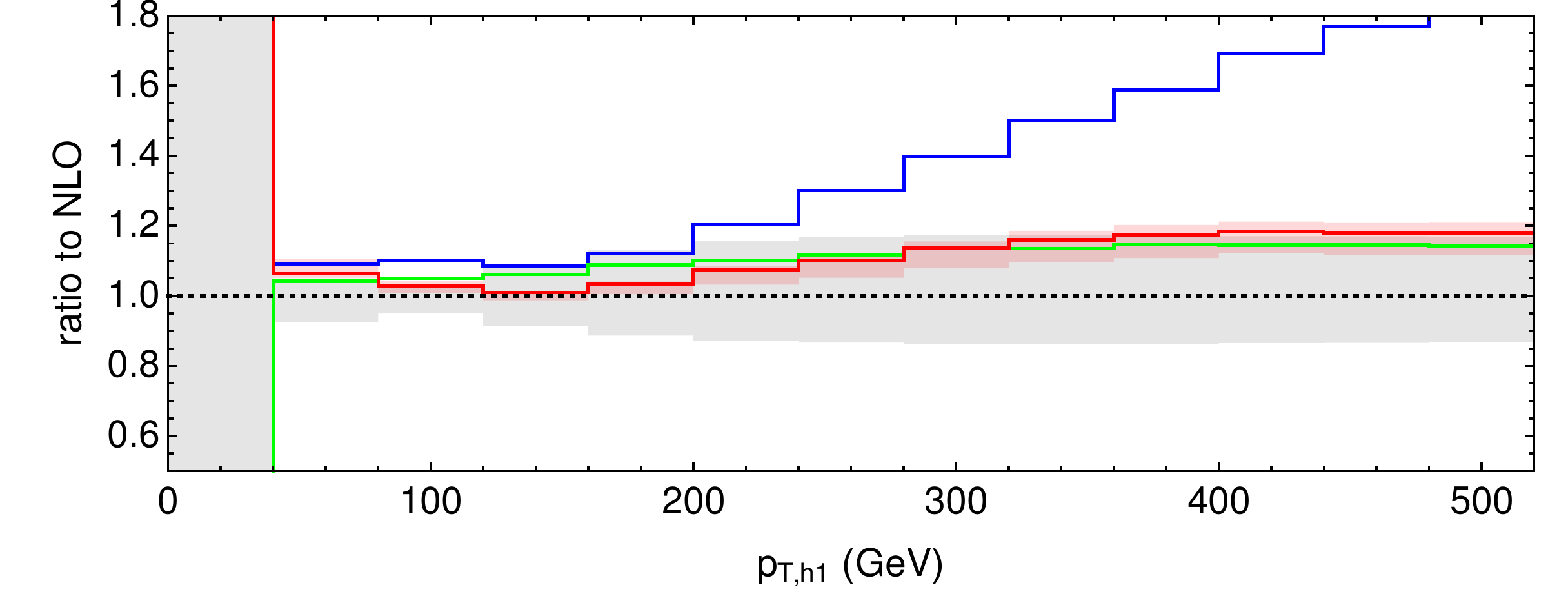}
\end{center}
\vspace{-2ex}
\caption{\label{fig:pT_h1}\small
Transverse momentum distribution for the harder Higgs boson at $14\,$TeV (left) and $100\,$TeV (right).
}
\end{figure}
%
\begin{figure}[t!]
\begin{center}
\includegraphics[width=.48\textwidth]{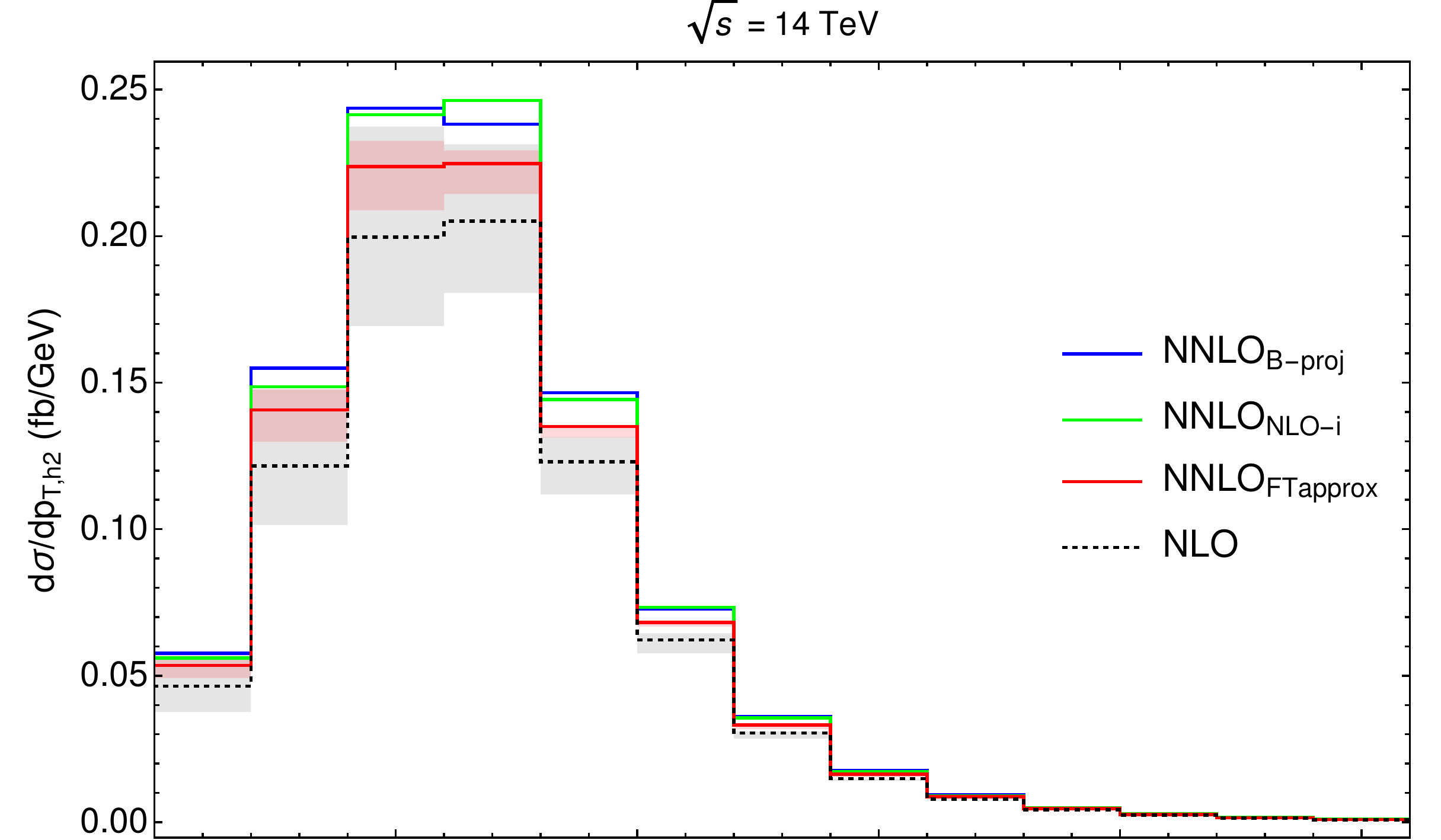}
\hfill
\includegraphics[width=.48\textwidth]{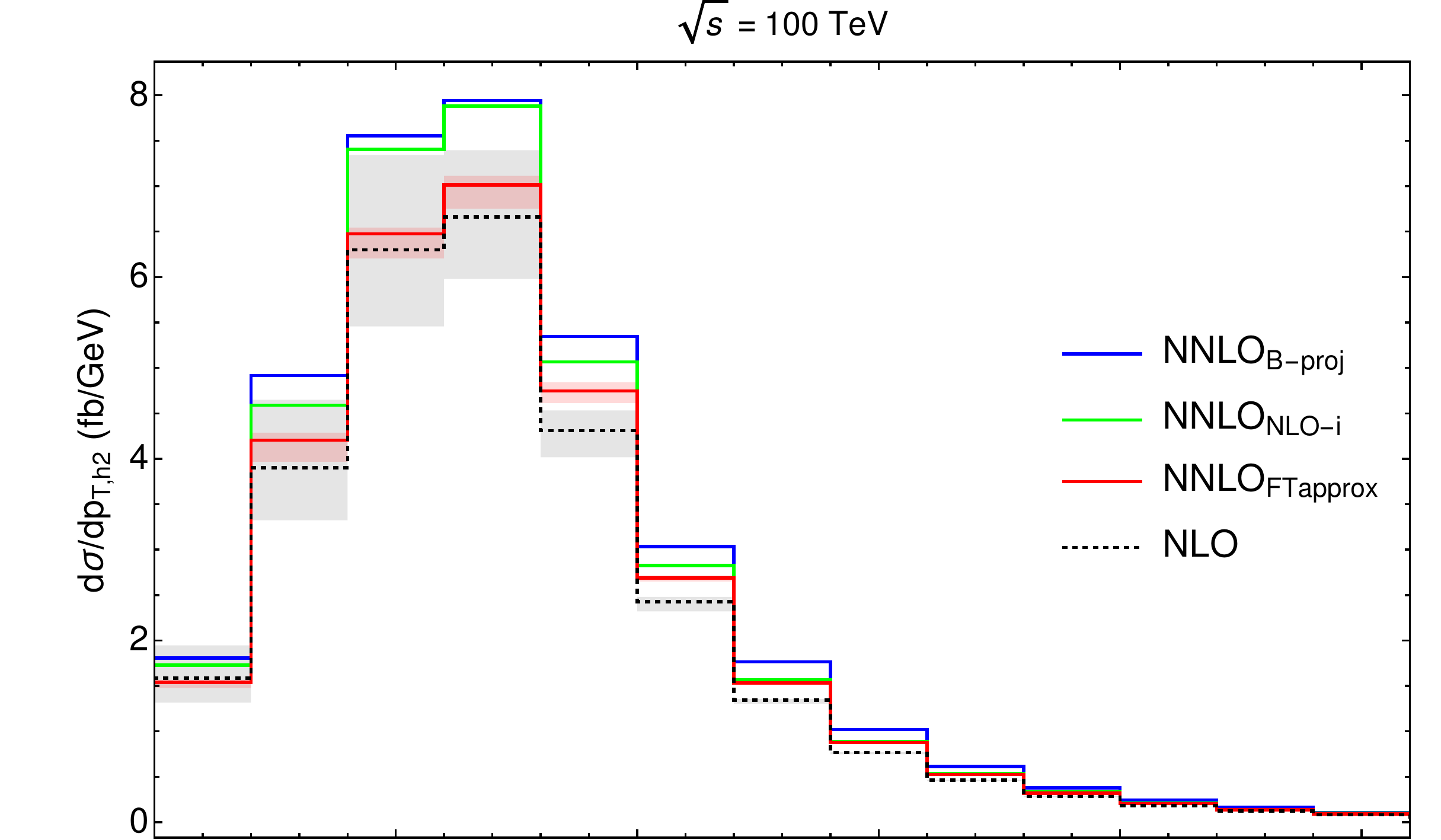}
\\
\includegraphics[width=.48\textwidth]{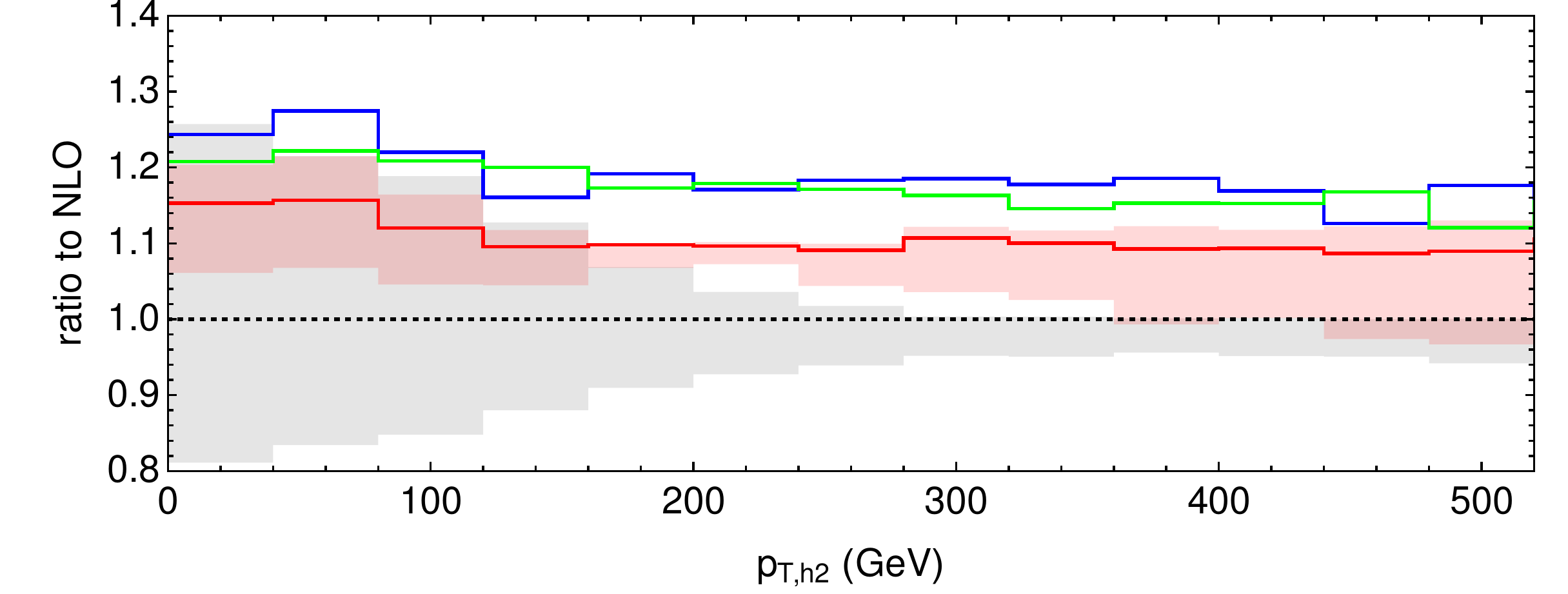}
\hfill
\includegraphics[width=.48\textwidth]{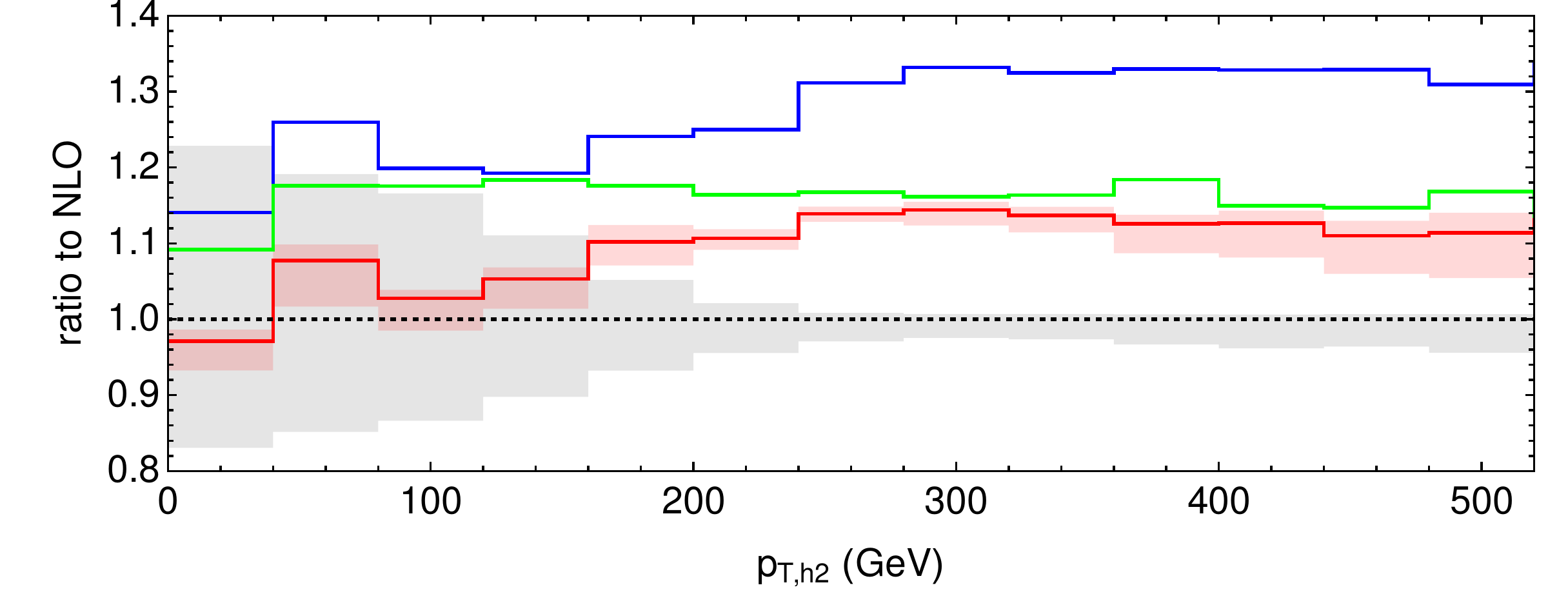}
\end{center}
\vspace{-2ex}
\caption{\label{fig:pT_h2}\small
Transverse momentum distribution for the softer Higgs boson at $14\,$TeV (left) and $100\,$TeV (right).
}
\end{figure}
As can be expected from the $p_{{\rm T},hh}$ spectrum, the NNLO$_\text{B-proj}$ result for $p_{{\rm T},h1}$ features very large corrections as $p_{{\rm T},h1}$ increases.
The effect, however, is less severe than the one observed in $p_{{\rm T},hh}$ because the $p_{{\rm T},h1}$ observable is already well defined at LO.
The NNLO$_\text{NLO-i}$ curve is overall in good agreement with the NNLO$_\text{FTapprox}$ prediction: It shows moderate corrections with respect to the NLO result which increase as $p_{{\rm T},h1}$ increases, while the scale uncertainties are about $\pm 15\%$. At very small $p_{{\rm T},h1}$ the higher-order corrections become perturbatively unstable as the available phase space for the real radiation is severely restricted in this regime yielding large logarithms that should be resummed in order to get a reliable prediction, see also the discussion in Section~3.4 of \citere{Heinrich:2017kxx}.
For the transverse momentum of the softer Higgs boson, $p_{{\rm T},h2}$, the NNLO effect is rather uniform in all three approximations, especially at $14\,$TeV.
The NNLO$_\text{FTapprox}$ predicts small corrections of order $10\%$, while the other two approximations show larger corrections with a similar shape.
In the tail of the distribution the scale uncertainty at NNLO is larger than at NLO, most likely due to an accidentally small size of the NLO scale variation (in fact, in this region the NLO corrections almost vanish).

\begin{figure}[t!]
\begin{center}
\includegraphics[width=.48\textwidth]{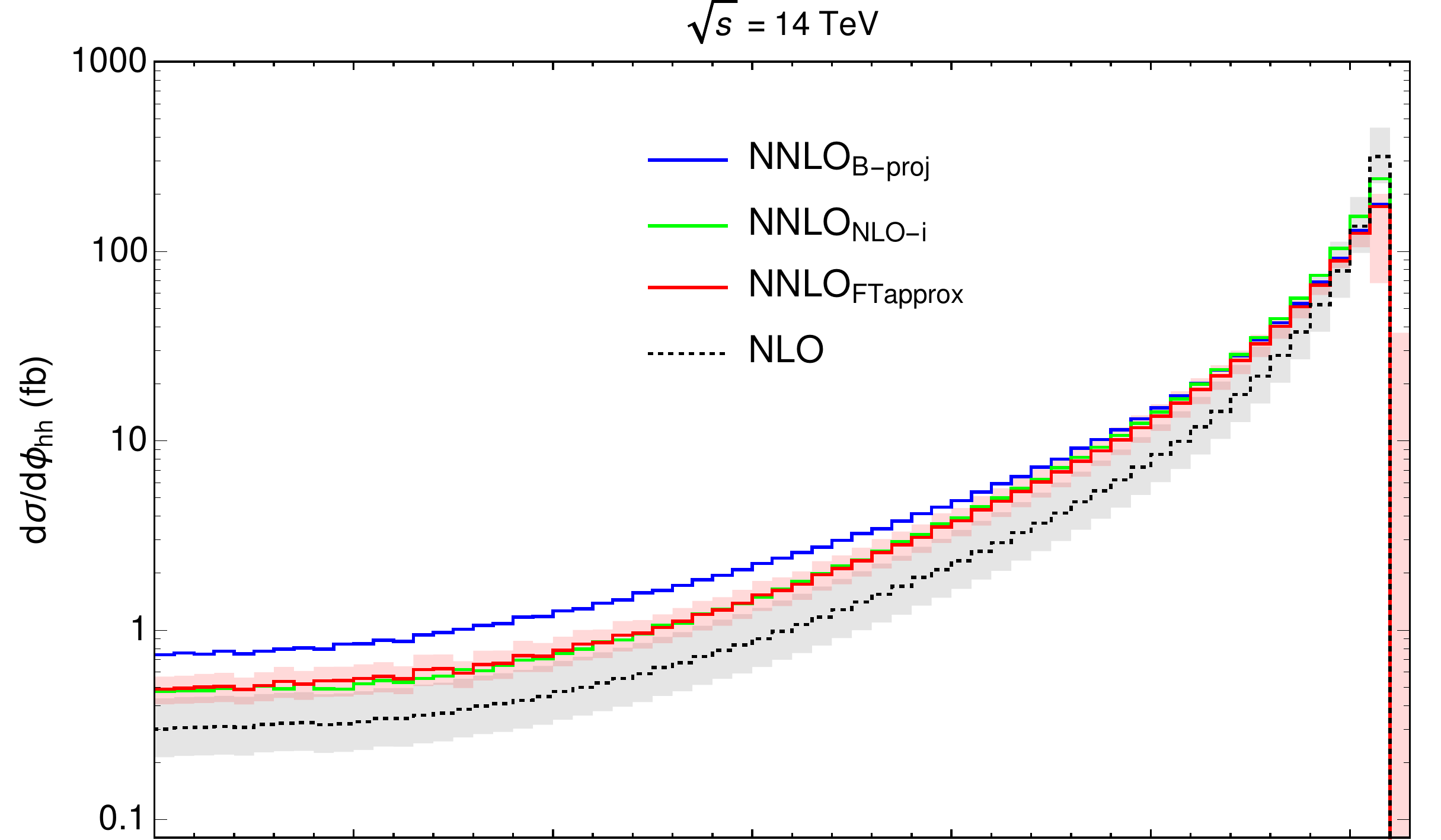}
\hfill
\includegraphics[width=.48\textwidth]{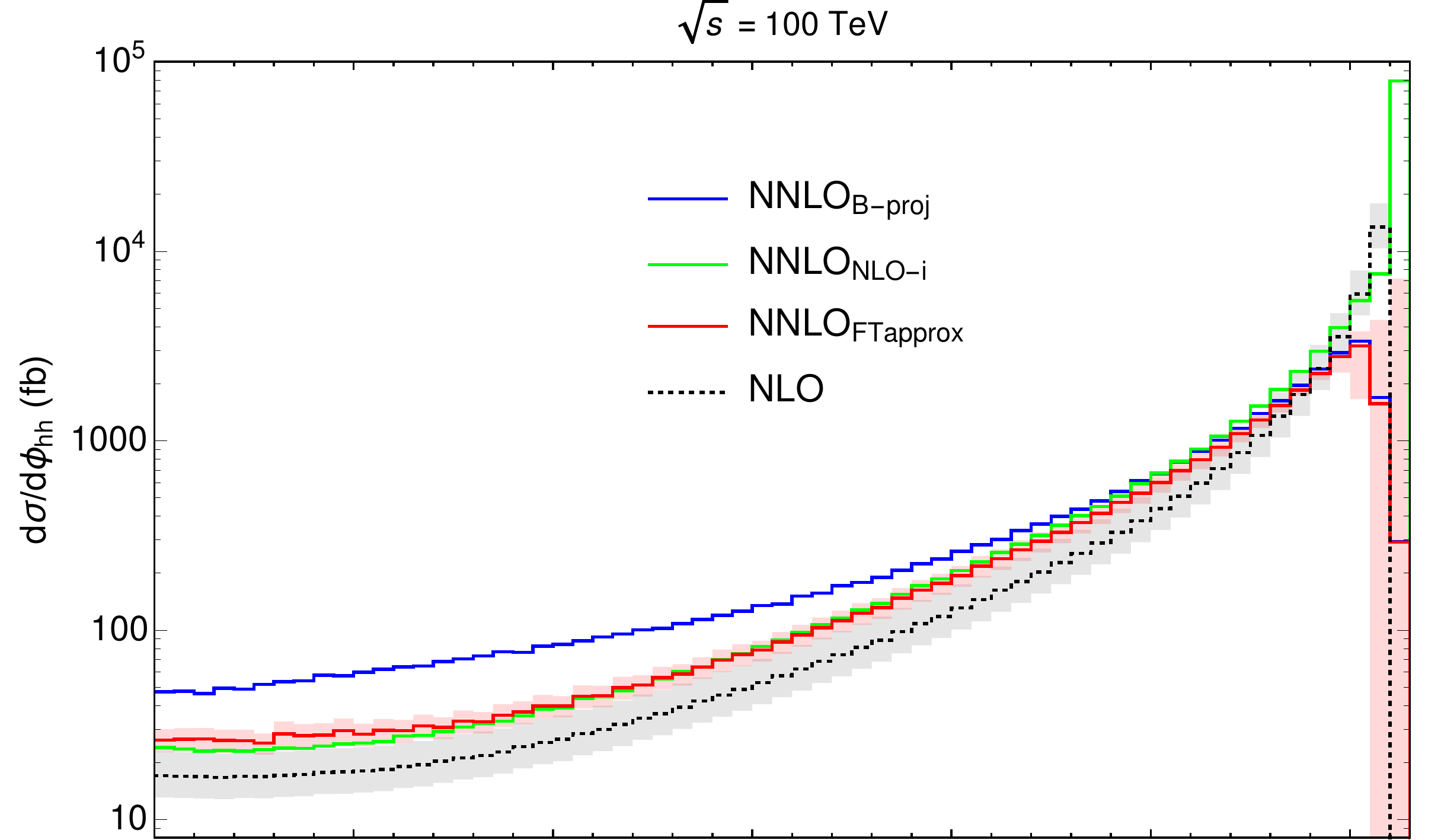}
\\
\includegraphics[width=.48\textwidth]{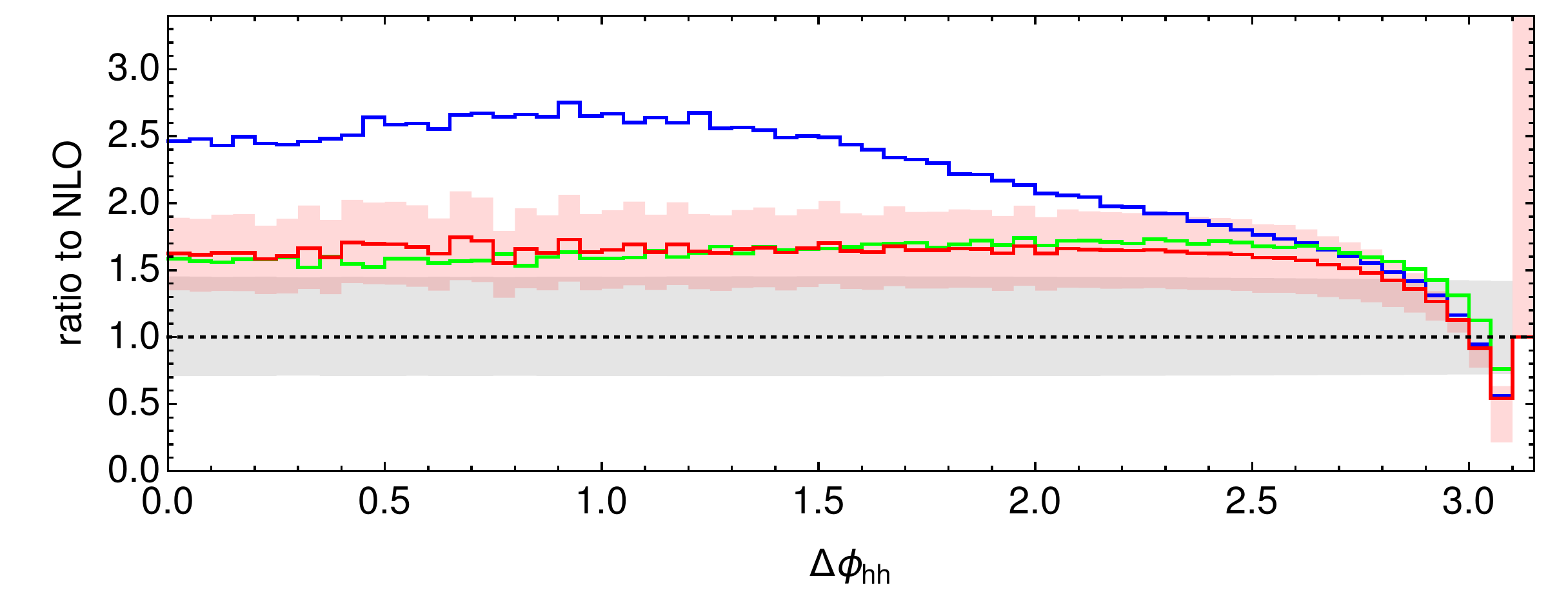}
\hfill
\includegraphics[width=.48\textwidth]{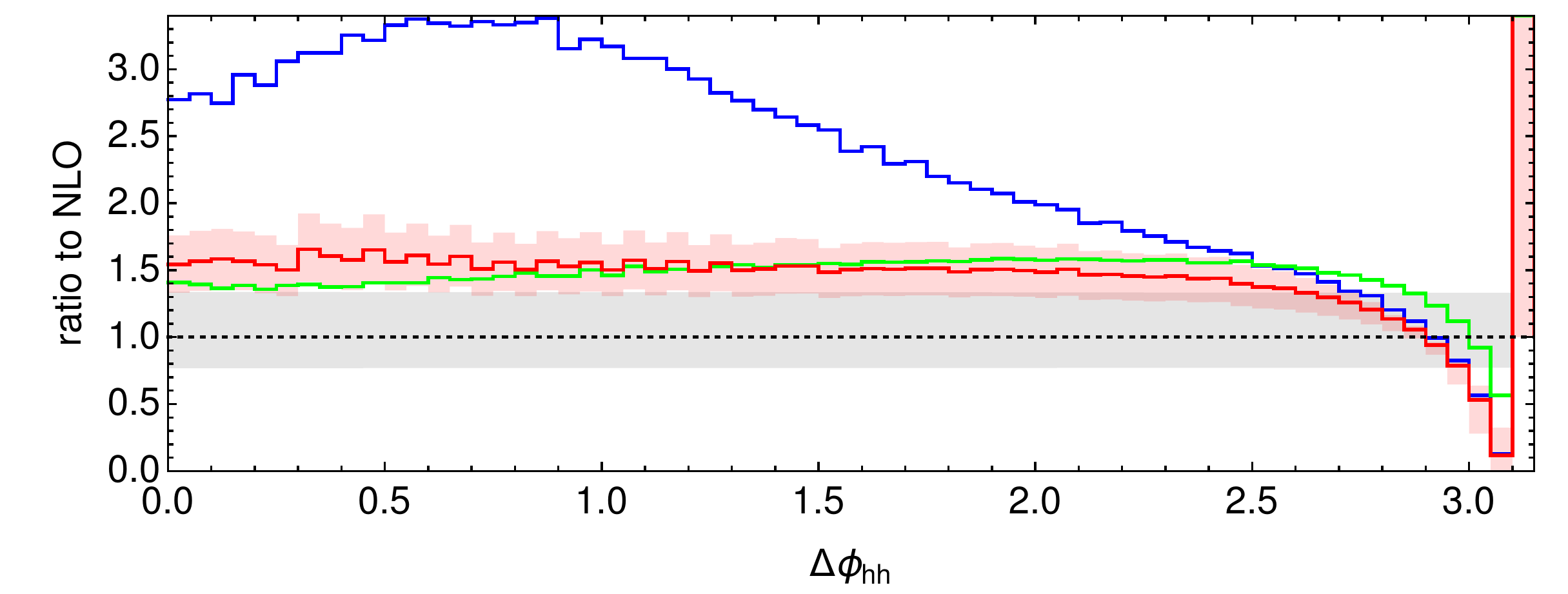}
\end{center}
\vspace{-2ex}
\caption{\label{fig:phi_hh}\small
Azimuthal angular separation between the two Higgs bosons at $14\,$TeV (left) and $100\,$TeV (right).}
\end{figure}

Finally, the distribution in the azimuthal angle between the two Higgs bosons, $\Delta\phi_{hh}$, is shown in \reffi{fig:phi_hh}. At LO we have $\Delta\phi_{hh} = \pi$, due to the back-to-back production of the two Higgs bosons at Born level. Real contributions allow $\Delta\phi_{hh}$ to be smaller than $\pi$, and again we observe that the NNLO$_\text{B-proj}$ approximation predicts larger corrections in the region dominated by hard radiation compared to the other two results, which again are in good agreement with each other in that region, whereas they start to deviate for larger angles. For values of $\Delta\phi_{hh}$ close to $\pi$, this observable receives large corrections from soft-gluon emission, and the corresponding large logarithms should be resummed in order to get a reliable prediction.


We conclude this section by adding a few comments on the finite-$M_t$ uncertainties at NNLO for the various differential distributions.
The analysis that was performed for the total cross section cannot be easily extended to differential distributions. On one hand, any accidental agreement between the \ftapprox and the full result at NLO in a given phase-space region would likely lead to an underestimation of the top quark mass effects; on the other hand, the regions in which the NLO corrections are very small due to cancellations between different contributions can present very large relative differences in the ${\cal O}(\as^3)$ contribution of the NLO$_\text{FTapprox}$ and NLO results, thus leading to artificially large uncertainties at NNLO.
In addition, there are observables that are by definition reproduced in an exact way by the \ftapprox at NLO (in our case $p_{{\rm T},hh}$, $p_{{\rm T},j1}$ and $\Delta\phi_{hh}$), and the uncertainty estimate procedure that we defined for the inclusive case is therefore not applicable. 
Despite these facts, and based on the performance of the \ftapprox at NLO~\cite{Borowka:2016ypz} as well as on the observed differences
between our NNLO approximations, we can try to assess the order of magnitude of the expected missing $M_t$ effects for the distributions presented above.

In the Higgs boson pair invariant-mass distribution, for values of $M_{hh}$ below $500\,$GeV the level of accuracy of the \ftapprox at NLO is similar to the inclusive case,
and therefore the $M_t$ uncertainty at NNLO is expected to be of a comparable size.
In the tail of the distribution, however, the quality of the \ftapprox decreases (see Fig.~5 of \citere{Borowka:2016ypz}),
and we thus expect the finite top quark mass effects to be of ${\cal O}(10\%)$ in this region.

The shape of the rapidity distribution of the Higgs boson pair is correctly described by the \ftapprox at NLO (see Fig.~8 of \citere{Borowka:2016ypz}), and the difference to the full result is only the overall normalization. Based on this, the estimated top quark mass uncertainty for the \nnloFT result is constant in the whole $y_{hh}$ range and of the same size as for the inclusive cross section.

The transverse momentum of the harder Higgs boson is very well described at NLO by the \ftapprox (see Fig.~7 of \citere{Borowka:2016ypz}), being always within the NLO scale uncertainty band. This fact, together with the close agreement between the \nnloFT and \nnloNI predictions, suggests that the missing top quark mass effects at NNLO are probably of moderate size.
The same holds true for the transverse-momentum distribution of the softer Higgs boson, except for the tail where at NLO the \ftapprox overestimates the full NLO corrections, which in fact almost vanish in this region.

The remaining distributions, which are either not defined or trivial at LO, are by definition reproduced in an exact way by the \ftapprox at NLO, and this makes the estimate of the missing top quark mass effects at NNLO more difficult. In this case, a possible approach can be to use the difference between the \nnloFT and \nnloNI prediction as an estimate of the uncertainty (as discussed before, the \nnloBP prediction is not expected to be reliable in the regions dominated by hard real radiation, where it largely deviates from the other two approximations).
This procedure would imply relatively low top quark mass uncertainties for the $p_{{\rm T},hh}$ and $\Delta\phi_{hh}$ distributions, except for the low $p_{{\rm T},hh}$ and the $\Delta\phi_{hh} \sim \pi$ regions, typically below the size of the scale uncertainties, and larger uncertainties for the leading-jet transverse momentum, for which the difference between the two approximations is larger.

%% file: conclusions.tex
In this work we considered Higgs boson pair production through gluon fusion in proton collisions.
We presented new QCD predictions for inclusive and differential cross sections, which include the full NLO contribution and also account for finite top quark mass effects at NNLO.
Our best prediction, denoted \nnloFT, retains the full top quark mass dependence in the double-real emission amplitudes, while the remaining real--virtual and two-loop virtual HEFT amplitudes are treated via a suitable reweighting for the corresponding subprocesses 
with a given final-state multiplicity.
This approximation represents the most advanced prediction available to date for this process.

The numerical results we obtained for the \nnloFT are quantitatively different from the results obtained in previous combinations.
In particular, as far as the total cross section is concerned,
the corrections turn out to be smaller than previous estimates, increasing the NLO result by about $12\%$ at $13\,$TeV and $7\%$ at $100\,$TeV.
The reduction of the scale uncertainties is significant, by about a factor of three for LHC energies.
Given that our \nnloFT prediction includes top quark mass effects in an approximated way, it is important to assess the corresponding uncertainty.
We carefully examined the performance of our approximations at both the inclusive and differential levels. The uncertainty on our reference inclusive \nnloFT prediction
is estimated to be about $\pm 2.7\%$ at $14\,$TeV, increasing with the collider energy to reach $\pm 4.6\%$ at $100\,$TeV.

Regarding differential distributions, in most of the cases we can observe clear qualitative differences with respect to the bin-by-bin reweighting procedure introduced in \citere{Borowka:2016ypz}, in the shape and/or the normalization. For some of the distributions, however, specifically the tails of the $p_{T,hh}$ and $p_{T,h1}$ 
spectra, both approximations are in very good agreement.
We discussed an estimate of the uncertainty associated with top quark mass effects at NNLO at the differential level, and we found that in most of the cases its magnitude is comparable to the size of the scale uncertainties, except for the tails of some distributions where the uncertainty from missing $M_t$ effects can be dominant.

%% file: HH_FTapp_main.bbl
\begin{thebibliography}{10}
\bibitem{munich}
{\sc Munich} is the abbreviation of ``MUlti-chaNnel Integrator at Swiss (CH)
  precision''---an automated parton level NLO generator by S.~Kallweit. In
  preparation.
\end{thebibliography}
